\documentclass[12pt, letterpaper]{article}
\usepackage[utf8]{inputenc}
\usepackage{latexsym}
\usepackage{amsmath}
\usepackage{amsfonts}
\usepackage{amssymb}

\usepackage{graphicx}
\usepackage{amssymb}

\begin{document}

  \title{Quantification of intrinsic quality of a principal dimension in correspondence analysis and taxicab correspondence analysis}
  \author{Vartan Choulakian, Universit\'{e} de Moncton, Canada \and email: vartan.choulakian@umoncton.ca}
  
  \date{August 2021}
  \maketitle

\begin{abstract}
Collins(2002, 2011) raised a number of issues with regards to correspondence analysis (CA),
such as: qualitative information in a CA map versus quantitative information in the relevant
contingency table; the interpretation of a CA map is difficult and its
relation with the \% of inertia (variance) explained. We tackle these issues
by considering CA and taxicab CA (TCA) as a stepwise Hotelling/Tucker
decomposition of the cross-covariance matrix of the row and column
categories into four quadrants. The contents of this essay are: First, we
review the notion of quality/quantity in multidimensional data analysis as
discussed by Benz\'{e}cri, who based his reflections on Aristotle. Second,
we show the importance of unravelling the interrelated concepts of
dependence/heterogeneity structure in a contingency table; and to picture
them two maps are needed. Third, we distinguish between intrinsic and
extrinsic quality of a principal dimension; the intrinsic quality is
based on the signs of the residuals in the four quadrants,
hence to the interpretability. Furthermore, we provide quantifications of
the intrinsic quality and use them to uncover structure in particular in sparse 
contingency tables. Finally, we emphasize the importance of looking at the
residual cross-covariance values at each iteration.


\end{abstract}

\textbf{In memoriam: Jean-Paul Benz\'{e}cri (28 February 1932, 24 November 2019)\bigskip }

\section{\textbf{Introduction}}

In this essay we comment on Collins' (2002, 2011) statement
\textquotedblright correspondence analysis makes you blind\textquotedblright
. Collins stated it twice as a reply to Whitlark and Smith (2001) and Bock
(2011a), who analyzed two different brand image count data sets by
correspondence analysis (CA). Both Whitlark and
Smith (2002) and Bock (2011b) in their reply to Collins (2002, 2011)
accepted without hesitation the insightful and original observations
described by Collins. Beh and Lombardo (2014, p.131-132) provide a cursory report on these dialogues.

{\tiny
\begin{tabular}{l|llllllll}
\multicolumn{9}{l}{\textbf{Table 1: WS brand-attribute count table.}} \\
\hline
\multicolumn{9}{c}{Attribute} \\
Company & innovative & leader & solution & rapport & efficient & relevant &
essential & trusted \\ \hline
Oracle & 155 & 157 & 109 & 133 & 151 & 96 & 35 & 170 \\
Nokia & 375 & 350 & 274 & 318 & 351 & 284 & 91 & 408 \\
Fedex & 476 & 675 & 550 & 669 & 748 & 627 & 307 & 754 \\ A & 86 & 66 & 105 & 110 & 117 & 76 & 30 & 122 \\
B & 30 & 21 & 25 & 37 & 40 & 20 & 9 & 43 \\
C & 18 & 12 & 11 & 16 & 17 & 12 & 2 & 18 \\
D & 25 & 23 & 33 & 36 & 34 & 28 & 12 & 35 \\
E & 21 & 20 & 21 & 26 & 27 & 18 & 9 & 36 \\
F & 190 & 307 & 305 & 332 & 355 & 309 & 131 & 392 \\
G & 18 & 16 & 16 & 25 & 21 & 18 & 10 & 29 \\
H & 408 & 549 & 467 & 551 & 613 & 523 & 239 & 624 \\
I & 143 & 225 & 194 & 191 & 206 & 184 & 121 & 248 \\ \hline
\end{tabular}%
}

\bigskip

{\tiny \ {%
\begin{tabular}{l||llllllll||l}
\multicolumn{10}{l}{\textbf{Table 2: Seriated WS brand-attribute data
structure uncovered by Collins.}} \\ \hline
\multicolumn{10}{c}{Attribute} \\
Company & trusted & efficient & rapport & leader & relevant & solution &
\multicolumn{1}{|l}{innovative} & essential & average \\ \hline
Fedex & 754 & 748 & 669 & 675 & 627 & 550 & 476 & 307 & 601 \\
H & 624 & 613 & 551 & 549 & 523 & 467 & 408 & 239 & 497 \\
Nokia & 408 & 351 & 318 & \textbf{350} & 284 & 274 & \textbf{375} & 91 & 306
\\
F & 392 & 355 & 332 & 307 & 309 & 305 & 190 & 131 & 290 \\
I & 248 & 206 & 191 & \textbf{225} & 184 & 194 & 143 & 121 & 189 \\
Oracle & 170 & 151 & 133 & \textbf{157} & 96 & 109 & \textbf{155} & 35 & 126
\\
A & 122 & 117 & 110 & 66 & 76 & \textbf{105} & 86 & 30 & 89 \\
D & 35 & 34 & 36 & 23 & 28 & \textbf{33} & 25 & 12 & 28 \\
B & 43 & 40 & 37 & 21 & 20 & 25 & \textbf{30} & 9 & 28 \\
E & 36 & 27 & 26 & 20 & 18 & 21 & 21 & 9 & 22 \\
G & 29 & 21 & 25 & 16 & 18 & 16 & 18 & 10 & 19 \\
C & 18 & 17 & 16 & 12 & 12 & 11 & \textbf{18} & 2 & 13 \\ \hline\hline
average & 240 & 223 & 204 & 202 & 183 & 176 & 162 & 83 & 184 \\ \hline
\end{tabular}%
}}

\bigskip

\subsection{WS brand-attribute count data}

Table 1, from Whitlark and Smith (2001), shows a brand-attribute count data.
Whitlark and Smith (2001) analyzed it by CA resulting in a map, Figure 1,
very similar to the Taxicab CA (TCA) map in Figure 2; they interpreted
Figure 1 using the adjusted chi-square residuals by arguing that it corrects
\textquotedblright inaccuracies introduced by the dimensionality
problem\textquotedblright. TCA is a \textit{l}%
$_{1}$ variant of CA, see Choulakian (2006).
The interpretation of Figures 1 and 2 does not
correspond to the simple seriated structure uncovered by Collins (2002)
given in Table 2, where the rows and columns of Table 1 are permuted
according to the row  and column averages in descending order. Collins
interpreted the seriated structure in Table 2 as:

\textquotedblright a relief map of a plain sloping steeply down from north (%
\textit{FedEx}) to south (\textit{Brand C}) and less steeply from west (%
\textit{Trusted}) to east (\textit{Essential}). On the plain would be a
molehill (\textit{Nokia} is more \textit{innovative}). But isn't the table
easy enough?\textquotedblright

Note that this simple structure uncovered by Collins is not found in Figures
1 and 2. Why? Lemma 1 provides a partial answer: The brands and the
attributes are not independent, because of the existence of \textit{multiple}
'molehill's, the most important molehill being (\textit{Nokia} is more
\textit{innovative}) as mentioned by Collins. The molehills (in bold in
Table 2) represent positive associations, and can be analyzed and visualized
in a complementary form in Figures (1 and 4) or in Figures (2 and 3).
Figures 3 and 4 are TCA and CA contribution plots, which help us to
interpret Figures 1 and 2. This is a summary of the paper.

\begin{figure}[]
    \begin{minipage}[c]{.46\linewidth}
        \centering
        \includegraphics[width=3.0in]{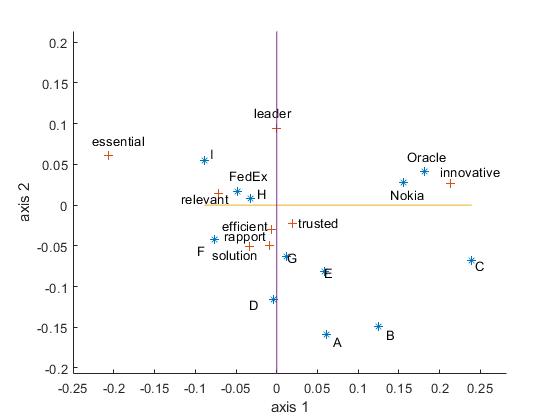}
        \caption{CA map of WS data.}
    \end{minipage}
    \hfill%
    \begin{minipage}[c]{.46\linewidth}
        \centering
        \includegraphics[width=3.0in]{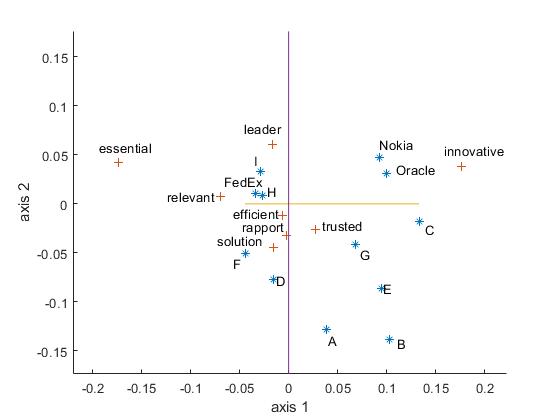}
        \caption{TCA map of WS data.}
    \end{minipage}
\end{figure}

\begin{figure}[]
    \begin{minipage}[c]{.46\linewidth}
        \centering
        \includegraphics[width=3.0in]{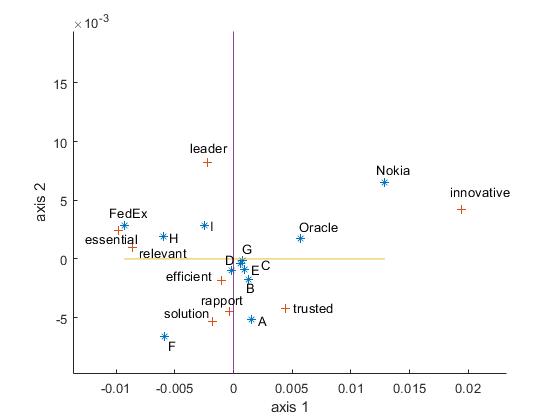}
        \caption{TCOV (TCA contribution map of WS data.}
    \end{minipage}
    \hfill%
    \begin{minipage}[c]{.46\linewidth}
        \centering
        \includegraphics[width=3.0in]{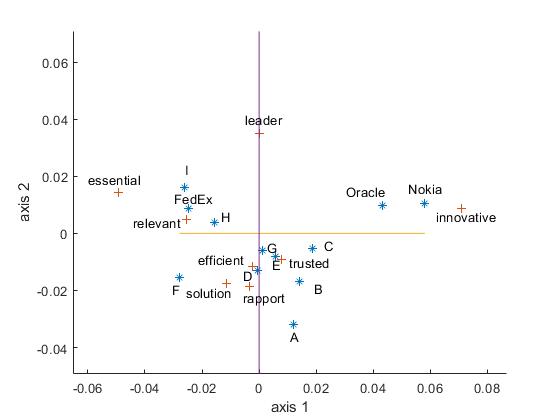}
        \caption{CA contribution map of WS data.}
    \end{minipage}
\end{figure}

\subsection{Issues raised by Collins}

Collins (2002) summarized his arguments against CA in Whitlark and Smith (2001) paper as:

\textquotedblright The popularity of CA and other techniques seems to arise
from two feelings: The data are too complex to be handled by the human
brain, and \textit{pictures communicate more than tables}. In fact, the
analysis of brand "image" data like that given by the authors is easy, and
\textit{quantitative patterns are best communicated through simple tables
supported by words}. \textit{A picture like the correspondence "map" shown
by W\&S may say something qualitative about patterns in the data, but it
says nothing quantitative or testable.} Not even, for example, \textit{that
the two dimensions of the map account for nearly 90\% of the variance of the
data}, as reported in the text.\textquotedblright

The highlighted parts are ours. There are two main points raised by Collins.

First, qualitative patterns in CA map Figure 1 do not reflect quantitative
patterns in Table 2. Friendly and Kwan (2011) divided statisticians (data
analysts) in two categories: graph-people and table-people. Clearly Collins
belongs to the group of table-people. In this paper, we will show that both,
tables and maps, are needed for global and local analysis. Additionally, in
section 2 we will discuss the important notions of quality and quantity in
data analysis, based on Benz\'{e}cri's reflections.

Second, the interpretation of a CA map is difficult and its relation with
the \% of inertia (variance) explained. Here, we have to discuss two
interrelated issues in CA. a) It concerns the interpretation of CA maps,
which as we mentioned, was raised by Whitlark and Smith (2001); then clearly
stated by Bock (2011b, pp. 587--588), and further discussed by Bock (2017)
in a R-Blog titled \textquotedblright\ How to Interpret Correspondence
Analysis Plots (It Probably Isn't the Way You Think)\textquotedblright ; in
the R-blog are discussed nine complex issues for the interpretation of CA
maps. Beh and Lombardo (2014, p.132) summarize it as: in CA
\textquotedblright the inability of the principal coordinate to provide a
meaningful interpretation of the distance between a row and column point in
these plots\textquotedblright. Greenacre (2013) also proposed contribution
biplots to tackle this issue. b) In this paper, we will distinguish between
the intrinsic quality and the extrinsic quality of a principal dimension, and
introduce indices that quantify the intrinsic quality of a principal
axis. In CA maps \textquotedblright variance accounted
for\textquotedblright\ reflects the extrinsic quality. The intrinsic
quality of a principal dimension examines the four quadrants of the
residual cross-covariance matrix. We tackle these issues by considering CA
and TCA as a stepwise Hotelling/Tucker decomposition of the residual
cross-covariance matrix of the row and column categories into four quadrants.

\subsection{Hotelling and Tucker decompositions}

In mathematics, a data set $\mathbf{X}=(x_{ij})$ for $i=1,...,I$ and $%
j=1,...,J$ can be interpreted as three kinds of mapping, see Benz\'{e}cri
(1973, p.56) and Choulakian (2016a). First, as a linear mapping: \textbf{X}:
R$^{J}\rightarrow $ R$^{I};$ second, as a linear mapping: \textbf{X}$^{T}$: R%
$^{I}\rightarrow $ R$^{J};$ third, as a bilinear mapping \textbf{X}: (R$%
^{I}, $ R$^{I})\rightarrow $ R. Hotelling's (1933) principal components
analysis (PCA) is developed within the first two settings, while
Hotelling's (1936) canonical correlation analysis is developed
within the third setting, where \textbf{X} represents a
cross-covariance matrix. Benz\'{e}cri (1973) emphasized the development of
CA within the first two settings as a weighted PCA method. In this paper, we
shall emphasize the third setting.

It is well known that CA is a particular kind of Hotelling's canonical
correlation analysis, see for instance Goodman (1991), where the two sets of
variables are the indicator sets of the categories of the two nominal
variables. Another method, similar in perspective to canonical correlation
analysis is Tucker's (1958) interbattery analysis, which maximizes the
covariance measure between the linear combination of the two indicator sets
of quantitative variables. When CA did not produce interpretable maps of
contingency tables, Tenenhaus and Augendre (1996) proposed the Tucker
interbattery analysis as an alternative to CA.

The parameters in Hotelling's canonical correlation and Tucker's covariance
analyses are generally estimated by singular value decomposition (SVD). When we
estimate the parameters by TaxicabSVD (TSVD) introduced by Choulakian (2006)
in place of SVD, surprisingly we notice that these two analyses complement
each other because they are linearly related, see equation (17). For further
details, see Choulakian, Simonetti and Gia (2014).

Figure 3, named TCov map, displays taxicab interbattery analysis map of WS data. Figures 2
and 3 (TCA and TCov maps) are different (similarly Figures 1 and 4 (CA and CA contribution maps) are different): thus they
provide different information to us, sometimes confusing (for instance
observe the positions of the brands \textit{B, E} and \textit{F} in Figures 2 and 3). One of
the major novelties of this paper is that, we interpret Figure 3, the
taxicab interbattery analysis TCov map, as TCA contribution map, and
consequently we provide a new perspective on the interpretation of the
associated TCA map Figure 2 via Lemma 6. For the interpretation of the row
and column labels on the TCov map, we shall use a quantification of the
intrinsic quality of a principal dimension, named quality of signs of
residuals (QSR) index; which will be complimented by a look at the seriated
residual covariance matrix. Then we extend the development of these ideas to
CA; where we also discuss sparse contingency tables having the quasi-two
blocks diagonal structure, which, according to Benz\'{e}cri (1973,
p.188-190), is quite common. Greenacre (2013) introduced and discussed CA
contribution biplots, but did not relate them to CA maps; Lemma 6 accomplishes this task.

\subsection{Organisation}

This paper is organized as follows: Section 2 sketches Benz\'{e}cri's
reflections on quality and quantity in data analysis. Section 3 presents an
overview of taxicab singular value decomposition (TSVD); section 4 presents
preliminaries. In section 5 we develop the main subject matter, the
quantification of the intrinsic quality of a principal dimension in CA and
TCA. Section 6 presents applications. Finally we conclude in section 7.

\section{Quantity and Quality}

Benz\'{e}cri (1982, 1988) has two papers on quality and quantity; in the
second he discussed the relationship between quantity and quality
historically, starting with Aristotle and finishing it with his description
within the philosophy of data analysis, aka CA framework. Here we quote from
Benz\'{e}cri (1988, section 1.7):

\textit{\textquotedblright Pour l'analyse des donn\'{e}es, nous retenons
d'abord, suivant Aristote, que \textquotedblright le caract\`{e}re propre de
la quantit\'{e} qu'on peut lui attribuer l'\'{e}gal et l'in\'{e}%
gal\textquotedblright , tandis que, \textquotedblright semblabe ou
dissemblable se dit uniquement des qualit\'{e}s\textquotedblright . De ce
point de vue, une description multidimensionnelle est toujours qualitative m%
\^{e}me si elle comporte que des variables num\'{e}riques pr\'{e}cises,
parceque la multiplicit\'{e} des descriptions possibles est telle qu'on
rencontrera jamais d'\'{e}gales, mais seulement de
semblables\textquotedblright .\bigskip }

The following two definitions and the corollary provide a succinct summary
of the quote.\textit{\bigskip }

\textbf{Definition 1 (Aristotle on quantity)}: X is a quantitative variable
if, given two realizations $x$ and $y$ of X, then either $x=y$ or $x\neq y.$

\textbf{Definition 2 (Aristotle on quality):} X is a qualitative variable
if, given two realizations $x$ and $y$ of X, then either $x$ is similar to $%
y $ or $x$ is dissimilar to $y.$

\textbf{Corollary 3 (Benz\'{e}cri)}: Any multidimensional description is
always qualitative even though its components are precisely numerical.$%
\bigskip $

Similar ideas also are expressed in a forward essay by Benz\'{e}cri in Murtagh (2005).

Benz\'{e}cri's schematic conceptual formulation of data analysis is the following directed diagram%
\begin{equation*}
\text{Quality(data)}\longrightarrow \text{Quantity(factors)}\longrightarrow \text{Quality(clusters).}
\end{equation*}

The first step: Quality$($data$)\longrightarrow \ \ $Quantity$($factors$)$
is done by dimension reduction. The nature of each factor (latent variable)
is quantitative and there are almost always more than one factor. Even
though each latent variable is quantitative, but its interpretation is
qualitative: According to Benz\'{e}cri (1988, section 1.3, in comments on
Descartes): \textquotedblright \textit{Toute qualit\'{e} n'est que
l'expression d'un rapport de quantit\'{e}s}\textquotedblright ; that is,
quality is the expression of a ratio of quantities. For interpretation of a
principal dimension, we apply \textit{Aristotle/Benz\'{e}cri principle}.
Aristotle in his book PHYSICS defined \textquotedblright \textit{principles
are contraries}\textquotedblright\ and cited as examples taken from his
predecessors \textquotedblright \textit{hot and cold}\textquotedblright,
\textquotedblright \textit{the rare and the dense}\textquotedblright\ and
\textquotedblright \textit{plenum and void}\textquotedblright\ see Aristotle
(1960, p.14). In CA, Benz\'{e}cri (1973, p.227) following Aristotle based
the interpretation of a principal dimension on contraries (dichotomies,
oppositions) and gradations, where an opposition or a gradation represent a
latent variable. In another context, Choulakian (2014, 2016b) used Euclid's
\textit{principle of contradiction} for interpretation of the first
principal dimension for the analysis of rank data.

The second step: Quantity$($factors$)\longrightarrow \ \ $Quality$($%
clusters) is done by usual methods such as k-means.

Murtagh (2005, section 1.1), described Benz\'{e}cri's paradigm
\textquotedblleft a tale of three metrics\textquotedblright ; which clearly
characterizes the diagram where : the chi-squared and the Euclidean metrics
are for the first step, and the ultrametrics for the second step. A similar
description to the above diagram is also stated by De Leeuw (2005).
This fact also is reflected in the first printed work, Benzécri (1973), titled \textit{DATA ANALYSIS};
 which is composed of two volumes: The first volume's subtitle is \textit{La Taxonomie};
 the second volume's subtitle is \textit{Analyse des Correspondances}.

\section{An overview of taxicab\ singular value decomposition}
Consider a matrix $\mathbf{X}$\ of size $I\times J$ and $rank(\mathbf{X})=k$%
\textbf{.} Taxicab singular value decomposition (TSVD) of \textbf{X} is a
decomposition similar to SVD of \textbf{X}; see Choulakian (2006, 2016a).

In TSVD the calculation of the dispersion measures $(\delta _{\alpha })$,
principal axes ($\mathbf{u}_{\alpha },\mathbf{v}_{\alpha })$ and principal
scores $(\mathbf{a}_{\alpha },\mathbf{b}_{\alpha })$ for $\alpha =1,...,k$
is done in an stepwise manner. We put $\mathbf{X}_{1}=\mathbf{X}=(x_{ij})$
and $\mathbf{X_{\alpha }}$ be the residual matrix at the $\alpha $-th iteration.

The variational definitions of the TSVD at the $\alpha $-th iteration are

\begin{eqnarray*}
\delta _{\alpha } &=&\max_{\mathbf{u\in
\mathbb{R}
}^{J}}\frac{\left\vert \left\vert \mathbf{X_{\alpha }u}\right\vert
\right\vert _{1}}{\left\vert \left\vert \mathbf{u}\right\vert \right\vert
_{\infty }}=\max_{\mathbf{v\in
\mathbb{R}
}^{I}}\ \frac{\left\vert \left\vert \mathbf{X_{\alpha }^{\prime }v}%
\right\vert \right\vert _{1}}{\left\vert \left\vert \mathbf{v}\right\vert
\right\vert _{\infty }}=\max_{\mathbf{u\in
\mathbb{R}
}^{J},\mathbf{v\in
\mathbb{R}
}^{I}}\frac{\mathbf{v}^{\prime }\textbf{X}\mathbf{_{\alpha }u}}{\left\vert \left\vert
\mathbf{u}\right\vert \right\vert _{\infty }\left\vert \left\vert \mathbf{v}%
\right\vert \right\vert _{\infty }}, \\
&=&\max ||\mathbf{X_{\alpha }u||}_{1}\ \ \text{subject to }\mathbf{u}\in
\left\{ -1,+1\right\} ^{J}, \\
&=&\max ||\mathbf{X_{\alpha }^{\prime }v||}_{1}\ \ \text{subject to }\mathbf{%
v}\in \left\{ -1,+1\right\} ^{I}, \\
&=&\max \mathbf{v}^{\prime }\mathbf{X_{\alpha }u}\text{ \ subject to \ }%
\mathbf{u}\in \left\{ -1,+1\right\} ^{J},\mathbf{v}\in \left\{ -1,+1\right\}
^{I}.
\end{eqnarray*}%
The $\alpha $-th principal axes are%
\begin{equation}
\mathbf{u}_{\alpha }\ =\arg \max_{\mathbf{u}\in \left\{ -1,+1\right\}
^{J}}\left\vert \left\vert \mathbf{X_{\alpha }u}\right\vert \right\vert _{1}%
\text{ \ \ and \ \ }\mathbf{v}_{\alpha }\ =\arg \max_{\mathbf{v}\in \left\{
-1,+1\right\} ^{I}}\left\vert \left\vert \mathbf{X_{\alpha }^{\prime }v}%
\right\vert \right\vert _{1}\text{,}
\end{equation}%
and the $\alpha $-th principal vectors are
\begin{equation}
\mathbf{a}_{\alpha }=\mathbf{X_{\alpha }u}_{\alpha }\text{ \ and \ }\mathbf{b%
}_{\alpha }=\mathbf{X_{\alpha }^{\prime }v}_{\alpha }.
\end{equation}%
Furthermore the following relations are also useful%
\begin{equation}
\mathbf{u}_{\alpha }=sgn(\mathbf{b}_{\alpha })\text{ \ and \ }\mathbf{v}%
_{\alpha }=sgn(\mathbf{a}_{\alpha }),
\end{equation}%
where $sgn(.)$ is the coordinatewise sign function, $sgn(x)=1$ \ if \ $x>0,$
\ and \ $sgn(x)=-1$ \ if \ $x\leq 0.$

The $\alpha $-th taxicab dispersion measure $\delta _{\alpha }$ can be
represented in many different ways

\begin{eqnarray}
\delta _{\alpha } &=&\mathbf{v}_{\alpha }^{\prime }\mathbf{X_{\alpha }u}%
_{\alpha } \\
&=&\left\vert \left\vert \mathbf{X_{\alpha }u}_{\alpha }\right\vert
\right\vert _{1}=\left\vert \left\vert \mathbf{a}_{\alpha }\right\vert
\right\vert _{1}=\mathbf{a}_{\alpha }^{\prime }\mathbf{v}_{\alpha },  \notag
\\
&=&\left\vert \left\vert \mathbf{X_{\alpha }^{\prime }v}_{\alpha
}\right\vert \right\vert _{1}=\left\vert \left\vert \mathbf{b}_{\alpha
}\right\vert \right\vert _{1}=\mathbf{b}_{\alpha }^{\prime }\mathbf{u}%
_{\alpha }.  \notag
\end{eqnarray}

The $(\alpha +1)$-th residual correspondence matrix is
\begin{equation}
\mathbf{X_{\alpha +1}}=\mathbf{X_{\alpha }-a}_{\alpha }\mathbf{b}_{\alpha
}^{\prime }/\delta _{\alpha }.
\end{equation}%
An interpretation of the term $\mathbf{a}_{\alpha }\mathbf{b}_{\alpha
}^{\prime }/\delta _{\alpha }$ in (5) is that, it represents the best rank-1
approximation of the residual correspondence matrix $\mathbf{X_{\alpha }}$,
in the sense of taxicab norm.

Thus TSVD of \textbf{X} corresponds to

\begin{equation}
x_{ij}=\sum_{\alpha =1}^{k}a_{\alpha }(i)b_{\alpha }(j)/\delta _{\alpha },
\end{equation}%
a decomposition similar to SVD, but where the vectors $(\mathbf{a}_{\alpha },%
\mathbf{b}_{\alpha })$ for $\alpha =1,...,k$ are conjugate, a weaker
property than orthogonality. That is%
\begin{equation*}
\mathbf{a}_{\alpha }^{\prime }sgn(\mathbf{a}_{\beta })=\mathbf{b}_{\alpha
}^{\prime }sgn(\mathbf{b}_{\beta })=0\text{ for }\alpha >\beta .
\end{equation*}

In TSVD, the calculation of the principal component weights, $\mathbf{u}%
_{\alpha }$ and $\mathbf{v}_{\alpha },$ and the principal scores, $\mathbf{a}%
_{\alpha }$\ and \ $\mathbf{b}_{\alpha },$ can be accomplished by two
algorithms. The first one is based on complete enumeration based on equation
(1). The second one is based on iterating the transition formulae (2,3).
This is an ascent algorithm; that is, it increases the value of the
objective function at each iteration, see Choulakian (2006, 2016a). The
iterative algorithm could converge to a local maximum; so it should be
restarted from several initial configurations. The rows or the columns of
the data can be used as starting values.

\section{Preliminaries}

Let $\mathbf{P=N/}n=(p_{ij})$ of size $I\times J$ be the associated
correspondence matrix of a contingency table \textbf{N}, where $%
n=\sum_{i=1}^{I}\sum_{j=1}^{J}N(i,j)$. We define as usual $p_{i\ast
}=\sum_{j=1}^{J}p_{ij}$ , $p_{\ast j}=\sum_{i=1}^{I}p_{ij},$ the vector $%
\mathbf{r=(}p_{i\ast })\in
\mathbb{R}
^{I},$ the vector $\mathbf{c=(}p_{\ast j})\in
\mathbb{R}
^{J}$, and $\mathbf{D}_{I}=Diag(\mathbf{r})$ the diagonal matrix having
diagonal elements $p_{i\ast },$ and similarly $\mathbf{D}_{J}=Diag(\mathbf{c}%
).$ We suppose that $\mathbf{D}_{I}$ and $\mathbf{D}_{J}$ are positive
definite metric matrices of size $I\times I$ and $J\times J$, respectively;
this means that the diagonal elements of $\mathbf{D}_{I}$ and $\mathbf{D}_{J}
$ are strictly positive. Let

\begin{eqnarray}
\mathbf{P}^{(1)} &=&(\mathbf{P}-\mathbf{rc}^{\top }) \\
&=&\mathbf{Cov}(\mathbf{P})  \notag
\end{eqnarray}
or elementwise
\begin{equation}
p_{ij}^{(1)}=p_{ij}-p_{i\ast }p_{\ast j}
\end{equation}%
be the residual matrix with respect to the independence model.\ $p_{ij}^{(1)}
$ is the cross-covariance between the categories of the $i$th nominal row
variable and the $j$th nominal column variable.

The independence assumption $p_{ij}^{(1)}=0$ can also be interpreted in
another way as

\begin{equation}
(\frac{p_{ij}}{p_{i\ast }p_{\ast j}}-1)=0,
\end{equation}%
which can be reexpressed as%
\begin{eqnarray}
\frac{1}{p_{i\ast }}(\frac{p_{ij}}{p_{\ast j}}-p_{i\ast }) &=&0 \\
&=&\frac{1}{p_{\ast j}}(\frac{p_{ij}}{p_{i\ast }}-p_{\ast j});  \notag
\end{eqnarray}%
this is the row and column homogeneity models. Benz\'{e}cri (1973, p.31)
named the vector ($\frac{p_{ij}}{p_{\ast j}}$ for $i=1,...,I$ and $j$ fixed)
the profile of the $j$th column; and the element $\frac{p_{ij}}{p_{i\ast
}p_{\ast j}}$ the density function of the probability measure $(p_{ij})$
with respect to the product measure $p_{i\ast }p_{\ast j}$. The element $%
\frac{p_{ij}}{p_{i\ast }p_{\ast j}}$is named Pearson ratio in Goodman (1996)
and Beh and Lombardo (2014, p.123).

\subsection{Estimation of the parameters by SVD}

Suppose the independence assumption $\mathbf{Cov(P)}=\mathbf{P}^{(1)}=%
\mathbf{0}$ is not true, then each of the two equivalent model formulations
(8,10) can be generalized to explain the nonindependence by adding bilinear
terms, where $k=rank(\mathbf{P}^{(1)}\mathbf{)}$.

a) Cov (cross-covariance) decomposition:%
\begin{equation}
p_{ij}-p_{i\ast }p_{\ast j}=\sum_{\alpha =1}^{k}a_{\alpha }(i)b_{\alpha
}(j)/\sigma _{\alpha }.
\end{equation}%
This is an interbattery analysis proposed by Tucker (1958). Tenenhaus and
Augendre (1996) estimated the parameters in (11) by singular value
decomposition (SVD) of the matrix $\mathbf{Cov(P).}$ The parameters in (11)
satisfy the following equations%
\begin{equation}
\sigma _{\alpha }^{2}=\sum_{i=1}^{I}|a_{\alpha
}(i)|^{2}=\sum_{j=1}^{J}|b_{\alpha }(j)|^{2}\text{\ \ \ for }\alpha =1,...,k;
\end{equation}

\begin{eqnarray}
0 &=&\sum_{i=1}^{I}a_{\alpha }(i)a_{\beta }(i)=\sum_{j=1}^{J}b_{\alpha
}(j)b_{\beta }(j)\text{\ \ for }\alpha \neq \beta  \\
&=&\sum_{i=1}^{I}a_{\alpha }(i)=\sum_{j=1}^{J}b_{\alpha }(j)\text{\ \ for }%
\alpha =1,...,k.  \notag
\end{eqnarray}

b) CA (correspondence analysis) decomposition%
\begin{equation}
\frac{p_{ij}}{p_{i\ast }p_{\ast j}}-1=\sum_{\alpha =1}^{k}f_{\alpha
}(i)g_{\alpha }(j)/\sigma _{\alpha }.
\end{equation}%
This decomposition has many interpretations. Essentially, for data analysis
purposes Benz\'{e}cri (1973) interpreted it as weighted principal components
analysis of row and column profiles. Another useful interpretation,
comparable to Tucker interbattery analysis, is Hotelling(1936)'s correlation
analysis, see Lancaster (1958) and Goodman (1991). The parameters in (14)
satisfy the following equations%
\begin{equation}
\sigma _{\alpha }^{2}=\sum_{i=1}^{I}|f_{\alpha }(i)|^{2}p_{i\ast
}=\sum_{j=1}^{J}|g_{\alpha }(j)|^{2}p_{\ast j}\text{\ \ \ for }\alpha
=1,...,k;
\end{equation}

\begin{eqnarray}
0 &=&\sum_{i=1}^{I}f_{\alpha }(i)f_{\beta }(i)p_{i\ast
}=\sum_{j=1}^{J}g_{\alpha }(j)g_{\beta }(j)p_{\ast j}\text{\ \ for }\alpha
\neq \beta \\
&=&\sum_{i=1}^{I}f_{\alpha }(i)p_{i\ast }=\sum_{j=1}^{J}g_{\alpha
}(j)p_{\ast j}\text{\ \ for }\alpha =1,...,k.  \notag
\end{eqnarray}

The above two decompositions given in (11) and (14) are cross-covariance
based. There are also association (log ratio) based decompositions see
Goodman (1991, 1996) or Greenacre and Lewi (2009).

\subsection{Estimation of the parameters by TSVD}

First, we estimate the parameters ($a_{\alpha }(i),\ b_{\alpha }(j),$ $%
\delta _{\alpha })$ in (11) by TSVD; then the parametrs in (14) will be
linearly related by
\begin{equation}
a_{\alpha }(i)=p_{i\ast }f_{\alpha }(i)\ ,\ \ b_{\alpha }(j)=p_{\ast
j}g_{\alpha }(j)\text{ and }\delta _{\alpha }=\sigma _{\alpha }.
\end{equation}%
The parameters $a_{\alpha }(i)$ and $b_{\alpha }(j)$ in (11) are the
principal coordinates of the TCov decomposition and they satisfy%
\begin{equation}
\delta _{\alpha }=\sum_{i=1}^{I}|a_{\alpha }(i)|=\sum_{j=1}^{J}|b_{\alpha
}(j)|\text{\ \ \ for }\alpha =1,...,k;
\end{equation}%
\begin{eqnarray*}
0 &=&\sum_{i=1}^{I}a_{\alpha }(i)sgn(a_{\beta }(i))=\sum_{j=1}^{J}b_{\alpha
}(j)sgn(b_{\beta }(j))\text{\ \ for }\alpha >\beta  \\
&=&\sum_{i=1}^{I}a_{\alpha }(i)=\sum_{j=1}^{J}b_{\alpha }(j)\text{\ \ for }%
\alpha =1,...,k.  \notag
\end{eqnarray*}%
Similarly, the parameters $f_{\alpha }(i)$ and $g_{\alpha }(j)$ in (14) are
the principal coordinates of the TCA decomposition and they satisfy%
\begin{equation}
\delta _{\alpha }=\sum_{i=1}^{I}|f_{\alpha }(i)|p_{i\ast
}=\sum_{j=1}^{J}|g_{\alpha }(j)|p_{\ast j}\text{\ \ \ for }\alpha =1,...,k;
\end{equation}%
\begin{eqnarray*}
0 &=&\sum_{i=1}^{I}f_{\alpha }(i)f_{\beta }(i)p_{i\ast
}=\sum_{j=1}^{J}g_{\alpha }(j)g_{\beta }(j)p_{\ast j}\text{\ \ for }\alpha
>\beta  \\
&=&\sum_{i=1}^{I}f_{\alpha }(i)p_{i\ast }=\sum_{j=1}^{J}g_{\alpha
}(j)p_{\ast j}\text{\ \ for }\alpha =1,...,k.
\end{eqnarray*}%
Let $\mathbf{P}^{(m)}=(p_{ij}^{(m)})$ be the $m$th residual correspondence
matrix, where%
\begin{equation}
p_{ij}^{(m+1)}=p_{ij}-p_{i\ast }p_{\ast j}-\sum_{\alpha =1}^{m}a_{\alpha
}(i)b_{\alpha }(j)/\delta _{\alpha }\text{\ \ \ for\ \ }m=1,...,k-1.
\end{equation}%
Similarly, let $\mathbf{D}^{(m)}=(d_{ij}^{(m)})$ be the $m$th residual
density matrix, where
\begin{equation}
d_{ij}^{(m+1)}=\frac{p_{ij}}{p_{i\ast }p_{\ast j}}-1-\sum_{\alpha
=1}^{m}f_{\alpha }(i)g_{\alpha }(j)/\delta _{\alpha }\text{\ \ \ for\ \ }%
m=1,...,k-1.
\end{equation}%
Let $S\cup \overline{S}=I$ be an optimal binary partition of $I$, and $T\cup
\overline{T}=J$ be an optimal binary partition of $J,$ such that $S=\left\{
i:a_{\alpha }(i)\geq 0\right\} $ and $T=\left\{ j:b_{\alpha }(j)\geq
0\right\} .$ Besides (18), the taxicab dispersion $\delta _{\alpha }$ will
additionally be related to the TCov principal coordinates $a_{\alpha }(i)$
and $b_{\alpha }(j)$ in (11) by the following useful equations:%
\begin{eqnarray}
\delta _{\alpha }/2 &=&\sum_{i\in S}a_{\alpha }(i)=-\sum_{i\in \overline{S}%
}a_{\alpha }(i) \\
&=&\sum_{j\in T}^{I}b_{\alpha }(j)=-\sum_{j\in \overline{T}}b_{\alpha }(j).
\notag
\end{eqnarray}

\begin{eqnarray}
\delta _{\alpha }/4 &=&\sum_{(i,j)\in S\times T}p_{ij}^{(\alpha
)}=\sum_{(i,j)\in \overline{S}\times \overline{T}}p_{ij}^{(\alpha )} \\
&=&-\sum_{(i,j)\in \overline{S}\times T}p_{ij}^{(\alpha )}=-\sum_{(i,j)\in
S\times \overline{T}}p_{ij}^{(\alpha )}.  \notag
\end{eqnarray}
Equations (22, 23) follow from the fact that $\mathbf{P}^{(\alpha )}$ for $%
\alpha =1,...,k$ is a double-centered matrix, see Choulakian and
Abou-Samra (2020). The quantification of the intrinsic quality of a
principal dimension is based on (23).

\subsection{An observation}

The TCov principal coordinates, $a_{\alpha }(i)$ and $b_{\alpha }(j),$ are
uniformly weighted, see equation (18); meanwhile TCA principal coordinates, $%
f_{\alpha }(i)$ and $g_{\alpha }(j),$ are marginally weighted, see equation
(19). What is the consequence to this? The answer to this question is: Benz%
\'{e}cri's principle of distributional equivalence, which states that CA
(and TCA) results are not changed if two proportional columns or rows are
merged into one. This has the practical consequence that the effective size
of sparse and large data sets can be smaller than the observed size; for
further details concerning sparse contingency tables see Choulakian (2017).

\section{Main developments}

Let \textbf{L} be a permutation matrix such that the coordinates of $\mathbf{%
r}_{L}=\mathbf{Lr}$ are in decreasing order, $\mathbf{r}_{L}(i)\geq \mathbf{r%
}_{L}(i+1)$ for $i=1,...,I-1.$ Similarly, \textbf{M} be a permutation matrix
such that the coordinates of $\mathbf{c}_{M}=\mathbf{Mc}$ are in decreasing
order, $\mathbf{c}_{M}(j)\geq \mathbf{c}_{M}(j+1)$ for $j=1,...,J-1.$

We consider the matrix
\begin{equation}
\mathbf{LP}^{(1)}\mathbf{M}^{\top }=\mathbf{L}(\mathbf{P}-\mathbf{rc}^{\top
})\mathbf{M}^{\top }.
\end{equation}%
We have the following easily proved result\bigskip

\textbf{Lemma 4}: Let $\mathbf{S=LPM}^{\top }.$ A \textit{necessary}
condition for the independence model, $\mathbf{LP}^{(1)}\mathbf{M}^{\top }=%
\mathbf{0}$ or $p_{ij}-p_{i\ast }p_{\ast j}=0,$ is that

\begin{equation}
\mathbf{S}(i,j)\geq \mathbf{S}(i,j+1)\text{ \ \ for\ \ \ }j=1,...,J-1\text{
and }i\text{ fixed}
\end{equation}%
and
\begin{equation}
\mathbf{S}(i,j)\geq \mathbf{S}(i+1,j)\text{ \ \ for\ \ \ }i=1,...,I-1\text{
and }j\text{ fixed.}
\end{equation}

\textbf{Remark}: Relations (25 and 26) characterize Robinson matrices used
for seriation of artifacts or sites in archeology. That is why we named $%
\mathbf{S,}$ see Table 2, seriated contingency table following its seriated
row $\mathbf{r}_{L}$ and column $\mathbf{c}_{M}$ marginals.\bigskip

\textbf{Lemma 5}: TSVD of $\mathbf{P}$ is equivalent to TCov($\mathbf{P})=$ TSVD of
$\mathbf{P}^{(1)}.$\bigskip

Lemma 6 states that the $\alpha $th row TCA (or CA) principal factor score $%
\mathbf{f}_{\alpha }(i)$ is the weighted covariance of the $\alpha $th
residual density function $\mathbf{D}^{(\alpha )}(i,:)$ with the $\alpha $%
-th principal axis $\mathbf{u}_{\alpha }$; where $\mathbf{D}^{(\alpha )}(i,:)$ is 
the ith row of $\mathbf{D}^{(\alpha )}$ and $\mathbf{D}^{(\alpha )}(:,j)$ is
the jth column of $\mathbf{D}^{(\alpha )}.$ 
\bigskip\

\textbf{Lemma 6}: In CA and TCA%
\begin{eqnarray}
\mathbf{f}_{\alpha }(i) &=&\mathbf{D}^{(\alpha )}(i,:)\mathbf{D}_{J}\mathbf{u%
}_{\alpha } \\
&=&cov(\mathbf{D}^{(\alpha )}(i,:),\mathbf{u}_{\alpha })  \notag
\end{eqnarray}%
and%
\begin{eqnarray}
\mathbf{g}_{\alpha }(j) &=&\mathbf{v}_{\alpha }^{\top }\mathbf{D}_{I}\mathbf{%
D}^{(\alpha )}(:,j) \\
&=&cov(\mathbf{D}^{(\alpha )}(:,j),\mathbf{v}_{\alpha }),  \notag
\end{eqnarray}%
where $\mathbf{u}_{\alpha }$ represents the $\alpha $-th standardized
principal axis in each method. In CA, $\mathbf{u}_{\alpha }=$ $\frac{\mathbf{%
g}_{\alpha }}{\sigma _{\alpha }}$and $\mathbf{v}_{\alpha }=\frac{\mathbf{f}%
_{\alpha }}{\sigma _{\alpha }}.$ In TCA, $\mathbf{u}_{\alpha }=sign(\mathbf{b%
}_{\alpha })=sign(\mathbf{g}_{\alpha })$ and $\mathbf{v}_{\alpha }=sign(%
\mathbf{a}_{\alpha })=sign(\mathbf{f}_{\alpha })$ for $\alpha =1,...,k$, see
equations (3 and 17).

Proof: Here, we provide a proof for TCA. We use the transition formula (2)
for\ \ $m=1,...,k-1,$
\begin{eqnarray*}
\mathbf{a}_{m+1}(i) &=&\sum_{j}p_{ij}^{(m+1)}\mathbf{u}_{m+1}(j) \\
&=&\sum_{j}\left[ p_{ij}-p_{i\ast }p_{\ast j}-\sum_{\alpha =1}^{m}a_{\alpha
}(i)b_{\alpha }(j)/\delta _{\alpha }\right] \mathbf{u}_{m+1}(j) \\
&=&\sum_{j}\left[ p_{ij}-p_{i\ast }p_{\ast j}-\sum_{\alpha =1}^{m}a_{\alpha
}(i)b_{\alpha }(j)/\delta _{\alpha }\right] \frac{p_{i\ast }p_{\ast j}}{%
p_{i\ast }p_{\ast j}}\mathbf{u}_{m+1}(j) \\
&=&\sum_{j}\left[ \frac{p_{ij}}{p_{i\ast }p_{\ast j}}-1-\sum_{\alpha
=1}^{m}f_{\alpha }(i)g_{\alpha }(j)/\delta _{\alpha }\right] p_{i\ast
}p_{\ast j}\mathbf{u}_{m+1}(j),\text{ by (17),} \\
\mathbf{f}_{m+1}(i) &=&\sum_{j}\left[ \frac{p_{ij}}{p_{i\ast }p_{\ast j}}%
-1-\sum_{\alpha =1}^{m}f_{\alpha }(i)g_{\alpha }(j)/\delta _{\alpha }\right]
p_{\ast j}\mathbf{u}_{m+1}(j) \\
&=&\sum_{j}d_{ij}^{(m+1)}p_{\ast j}\mathbf{u}_{m+1}(j),
\end{eqnarray*}%
which is the required result (27).

\textbf{Remark}: In CA, due to (16), equation (27), similarly (28), can
further be simplified to%
\begin{eqnarray*}
\mathbf{f}_{m+1}(i) &=&\sum_{j}\frac{p_{ij}}{p_{i\ast }p_{\ast j}}p_{\ast j}%
\mathbf{u}_{m+1}(j) \\
&=&cov(\mathbf{D}(i,:),\mathbf{u}_{m+1}),
\end{eqnarray*}%
a well known result in Bastin et. al. (1980, p.157) or Goodman(1991, p.
1105, eq. A.1.3).

\subsection{Quantifying the intrinsic quality of a taxicab principal axis}

Within the Euclidean framework a measure of the quality of a
principal dimension is the proportion (or percentage)  of the residual variance explained (or inertia in
the case of CA)%
\begin{equation*}
\%(\text{explained residual variance by dimension }\alpha )=\frac{100\ \sigma
_{\alpha }^{2}}{\sum_{\beta =\alpha}^{k}\sigma _{\beta }^{2}}.
\end{equation*}%
This is an extrinsic measure of quality, because it compares the
dispersion of a principal axis $\sigma _{\alpha }^{2}$ with the residual
dispersion $\sum_{\beta =\alpha}^{k}\sigma _{\beta }^{2}$. In the above
equation replacing the \textit{l}$_{2}$ terms by the corresponding \textit{l}%
$_{1}$ terms, we obtain the measure of intrinsic quality $QSR_{\alpha }$
expressed in Definition 7.

Let $S\cup \overline{S}=I$ be an optimal binary partition of $I$, and
similarly $T\cup \overline{T}=J$ be an optimal binary partition of $J$ for
the $\alpha $th principal dimension$.$ Thus the data set is divided into
four quadrants. We define a new index showing the quality of signs of the
residuals (QSR) in each quadrant of the $\alpha $th residual
cross-covariance matrix $\mathbf{P}^{(\alpha )}$ for $\alpha =1,...,k$ in
(20)$.$\bigskip

\textbf{Definition 7}:\ For $\alpha =1,...,k,$ the measure of the quality of
signs of the residuals in the quadrant $E\times F\subseteq I\times J$ is
\begin{eqnarray*}
QSR_{\alpha }(E,F) &=&\frac{\sum_{(i,j)\in E\times F}p_{ij}^{(\alpha )}}{%
\sum_{(i,j)\in E\times F}|p_{ij}^{(\alpha )}|},\text{ \ and by (23)} \\
&=&\frac{\delta _{\alpha }/4}{\sum_{(i,j)\in E\times F}|p_{ij}^{(\alpha )}|}%
\text{ \ \ for }(E,F)=(S,T)\text{ and (}\overline{S},\overline{T}) \\
&=&\frac{-\delta _{\alpha }/4}{\sum_{(i,j)\in E\times F}|p_{ij}^{(\alpha )}|}%
\text{ \ \ for }(E,F)=(\overline{S},T)\text{ and (}S,\overline{T}).
\end{eqnarray*}%
Similarly, a quantification of the quality of signs of the optimal cut of
dimension $\alpha $ is%
\begin{equation*}
QSR_{\alpha }=\frac{\delta _{\alpha }}{\sum_{(i,j)}|p_{ij}^{(\alpha )}|}%
\text{.}
\end{equation*}

\textbf{Remark: }The computation of the elements of $QSR_{\alpha }(E,F)$ are
done easily in the following way. We note that the $\alpha $th principal
axis can be written as%
\begin{equation*}
\mathbf{u}_{\alpha }=\mathbf{u}_{\alpha +}+\mathbf{u}_{\alpha -}\text{\ \ \
\ by\ \ (1)},
\end{equation*}%
where $\mathbf{u}_{\alpha +}=(\mathbf{u}_{\alpha }+\mathbf{1}_{J})/2\in
\left\{ 0,1\right\} ^{J}$ and $\mathbf{u}_{\alpha -}=(\mathbf{u}_{\alpha }-%
\mathbf{1}_{J})/2\in \left\{ -1,0\right\} ^{J};$ similarly
\begin{equation*}
\mathbf{v}_{\alpha }=\mathbf{v}_{\alpha +}+\mathbf{v}_{\alpha -}\text{\ \ \
\ \ by \ (1)},
\end{equation*}%
where $\mathbf{v}_{\alpha +}=(\mathbf{v}_{\alpha }+\mathbf{1}_{I})/2\in
\left\{ 0,1\right\} ^{I}$ and $\mathbf{v}_{\alpha -}=(\mathbf{v}_{\alpha }-%
\mathbf{1}_{I})/2\in \left\{ -1,0\right\} ^{I}$, and $\mathbf{1}_{I}$
designates a column vector of 1's of size $I.$ So%
\begin{eqnarray*}
QSR_{\alpha }(S,T) &=&QSR_{\alpha }(\mathbf{v}_{\alpha +},\mathbf{u}_{\alpha
+}) \\
&=&\frac{\delta _{\alpha }/4}{\mathbf{v}_{\alpha +}^{\prime }abs(\mathbf{X}%
_{\alpha })\mathbf{u}_{\alpha +}}>0,
\end{eqnarray*}%
\begin{eqnarray*}
QSR_{\alpha }(\overline{S},\overline{T}) &=&QSR_{\alpha }(\mathbf{v}_{\alpha
-},\mathbf{u}_{\alpha -}) \\
&=&\frac{\delta _{\alpha }/4}{\mathbf{v}_{\alpha -}^{\prime }abs(\mathbf{X}%
_{\alpha })\mathbf{u}_{\alpha -}}>0,
\end{eqnarray*}%
\begin{eqnarray*}
QSR_{\alpha }(S,\overline{T}) &=&QSR_{\alpha }(\mathbf{v}_{\alpha +},\mathbf{%
u}_{\alpha -}) \\
&=&\frac{\delta _{\alpha }/4}{\mathbf{v}_{\alpha -}^{\prime }abs(\mathbf{X}%
_{\alpha })\mathbf{u}_{\alpha +}}<0,
\end{eqnarray*}%
\begin{eqnarray*}
QSR_{\alpha }(\overline{S},T) &=&QSR_{\alpha }(\mathbf{v}_{\alpha -},\mathbf{%
u}_{\alpha +}) \\
&=&\frac{\delta _{\alpha }/4}{\mathbf{v}_{\alpha +}^{\prime }abs(\mathbf{X}%
_{\alpha })\mathbf{u}_{\alpha -}}<0,
\end{eqnarray*}%
where $abs(\mathbf{X}_{\alpha })=(|X_{\alpha }(i,j)|).$

To interpret the above indices, we recall from elementary probability theory
the definition of association between two events by defining an index of
association $ass(i,j)=p_{ij}-p_{i\ast }p_{\ast j}$ for $i=1,...,I$ and $%
j=1,...,J$.

a) When $ass(i,j)=0$, then the $i$th category of the row variable and the $j$%
th category of the column variable are \textit{not associated (independent).}

b) When $ass(i,j)>0$, then the $i$th category of the row variable and the $j$%
th category of the column variable are \textit{attractively or positively associated};
that is, the event (i,j) occurs \textit{more} than by chance.

c) When $ass(i,j)<0$, then the $i$th category of the row variable and the $j$%
th category of the column variable are \textit{repulsively or negatively associated; }that
is, the event (i,j) occurs \textit{less} than by chance.

Based on these, the interpretation of the indices becomes evident: for
instance, $QSR_{\alpha }(S,T)>0$ measures the intensity of the \textit{%
attractive association} between the subsets $S$ and $T;$ while $QSR_{\alpha
}(\overline{S},T)<0$ measures the intensity of the \textit{repulsive
association} between the subsets $\overline{S}$ $\ $\ and $\ T.$

Allard et al. (2020) used the QSR index to choose between two competing
methods of data analysis, TCA and taxicab log-ratio analysis of contingency
tables and compositional data.\bigskip

\textbf{Notation}:

$QSR_{\alpha }(+)=\left\{ QSR_{\alpha }(\mathbf{u}_{\alpha +},\mathbf{v}%
_{\alpha +}),QSR_{\alpha }(\mathbf{u}_{\alpha -},\mathbf{v}_{\alpha
-})\right\} $

$QSR_{\alpha }(-)=\left\{ QSR_{\alpha }(\mathbf{u}_{\alpha +},\mathbf{v}%
_{\alpha -}),QSR_{\alpha }(\mathbf{u}_{\alpha -},\mathbf{v}_{\alpha
+})\right\} \bigskip $

We have the following easily proved result\bigskip

\textbf{Lemma 8}: a) For $\alpha =1,...,k,$ $QSR_{\alpha }=1$ if and only if
$QSR_{\alpha }(S,T)=-QSR_{\alpha }(S,\overline{T})=-QSR_{\alpha }(\overline{S%
},T)=QSR_{\alpha }(\overline{S},\overline{T})=1.$

b) For $\alpha =k,$ $QSR_{\alpha }=1.$

c) ($QSR_{\alpha }(S,T)+|QSR_{\alpha }(S,\overline{T})|+|QSR_{\alpha }(%
\overline{S},T)|+QSR_{\alpha }(\overline{S},\overline{T}))/4\geq QSR_{\alpha
}.$

$\ \ \ $The proof of part c, is based on the arithmetic-harmonic means
inequality which states that for four strictly positive real numbers $a,b,c$
and $d$%
\begin{equation*}
\frac{a+b+c+d}{4}\geq \frac{4}{\frac{1}{a}+\frac{1}{b}+\frac{1}{c}+\frac{1}{d%
}};
\end{equation*}%
equality is attained when $a=b=c=d$.

\subsection{Quantifying the intrinsic quality of a principal axis in CA}

Let $\mathbf{D}^{(m)}=(d_{ij}^{(m)})$ be the $m$th residual density matrix
in CA,
\begin{eqnarray}
d_{ij}^{(m+1)} &=&\frac{p_{ij}}{p_{i\ast }p_{\ast j}}-1-\sum_{\alpha
=1}^{m}f_{\alpha }(i)g_{\alpha }(j)/\sigma _{\alpha } \\
&=&\frac{p_{ij}}{p_{i\ast }p_{\ast j}}-1-\sum_{\alpha =1}^{m}\sigma _{\alpha
}v_{\alpha }(i)u_{\alpha }(j)\text{\ \ \ for\ \ }m=1,...,k-1,  \notag
\end{eqnarray}%
where $\mathbf{u}_{\alpha }=\frac{\mathbf{g}_{\alpha }}{\sigma _{\alpha }}$
and $\mathbf{v}_{\alpha }=\frac{\mathbf{f}_{\alpha }}{\sigma _{\alpha }}$
represent the $\alpha $th standardized principal axis coordinates in CA.

Let $\mathbf{Q}^{(m)}=(q_{ij}^{(m)})$ be the $m$th residual cross-covariance
matrix in CA obtained from (29),
\begin{eqnarray}
q_{ij}^{(m+1)} &=&p_{i\ast }p_{\ast j}d_{ij}^{(m+1)} \\
&=&p_{ij}-p_{i\ast }p_{\ast j}-\sum_{\alpha =1}^{m}\sigma _{\alpha }p_{i\ast
}p_{\ast j}v_{\alpha }(i)u_{\alpha }(j)\text{\ \ \ for\ \ }m=1,...,k-1.
\notag
\end{eqnarray}

\subsubsection{$CA\_QSR$ $\ \ \ indices$}

Let $S\cup \overline{S}=I$ be an optimal binary principal axis partition of $%
I$, and similarly $T\cup \overline{T}=J$ be an optimal principal axis
partition of $J$ by CA. Thus the residual covariance matrix is divided into
four quadrants: $S=\left\{ i:v_{\alpha }(i)\geq 0\right\} $ and $T=\left\{
j:u_{\alpha }(j)\geq 0\right\} .$ Based on the observation that both $%
q_{ij}^{(m+1)}$ and $p_{ij}^{(m+1)}$ are double centered, we can quantify
the intrinsic quality of CA principal dimension by replacing $p_{ij}^{(m+1)}$
by $q_{ij}^{(m+1)}$ in subsection 5.1, and obtain $CA\_QSR$
measures.\bigskip

\textbf{Definition 9}:\ For $\alpha =1,...,k,$ the CA measure of the quality
of signs of the residuals in the quadrant $E\times F\subseteq I\times J$ is
\begin{eqnarray*}
CA\_QSR_{\alpha }(E,F) &=&\frac{\sum_{(i,j)\in E\times F}q_{ij}^{(\alpha )}}{%
\sum_{(i,j)\in E\times F}|q_{ij}^{(\alpha )}|} \\
&=&\frac{\varpi _{\alpha }/4}{\sum_{(i,j)\in E\times F}|q_{ij}^{(\alpha )}|}%
\text{ \ \ for }(E,F)=(S,T)\text{ and (}\overline{S},\overline{T}) \\
&=&\frac{-\varpi _{\alpha }/4}{\sum_{(i,j)\in E\times F}|p_{ij}^{(\alpha )}|}%
\text{ \ \ for }(E,F)=(\overline{S},T)\text{ and (}S,\overline{T}).
\end{eqnarray*}%
for $E=S$ and $\overline{S},$ and, $F=T$ and $\overline{T}$. Similarly the
CA measure of the quality of signs of principal dimension $\alpha $ is%
\begin{equation*}
CA\_QSR_{\alpha }=\frac{\varpi _{\alpha }}{\sum_{(i,j)}|q_{ij}^{(\alpha )}|},
\end{equation*}%
where%
\begin{eqnarray}
\varpi _{\alpha } &=&sign(\mathbf{f}_{\alpha }^{\prime })\ \mathbf{Q}%
^{(\alpha )}\ sign(\mathbf{g}_{\alpha }) \\
&=&\sum_{i=1}^{I}\sum_{j=1}^{J}sign(f_{\alpha }(i))sign(g_{\alpha }(j))\
q_{ij}^{(\alpha )}  \notag \\
&=&4\sum_{(i,j)\in E\times F}sign(f_{\alpha }(i))sign(g_{\alpha }(j))\
q_{ij}^{(\alpha )}\text{,}  \notag
\end{eqnarray}%
for $E=S$ and $\overline{S},$ and, $F=T$ and $\overline{T}.$

Note that equations (31) and (23) are similar, and they follow from the
important observation that both residual cross-covariance matrices $%
q_{ij}^{(m)}$ and $p_{ij}^{(m)}$ for $m=1,...,k$ are double centered.
Furthermore, the $CA\_QSR$ indices satisfy the three properties in Lemma 8.

We have the following \bigskip

\textbf{Lemma 10}: a) When both CA and TCA produce the same binary partition
(cut) of the set of column and the set of row variables of the cross-covariance
matrix $\mathbf{P}^{(1)}$, then $CA\_QSR_{1}=QSR_{1}$ $\ \ $and$\ \ \
CA\_QSR_{1}(E,F)=QSR_{1}(E,F)$ for $E=S$ and $\overline{S},$ and, $F=T$ and $%
\overline{T}$. Thus $\delta _{1}=\varpi _{1}.$

The proof is evident based on the fact that $p_{ij}^{(1)}=$ $%
q_{ij}^{(1)}=p_{ij}-p_{i\ast }p_{\ast j}.$ Note that for $\alpha \neq 1,$ $%
p_{ij}^{(\alpha )}\neq $ $q_{ij}^{(\alpha )}.$

b) When CA and TCA produce different binary partitions of the set of column
or the set of row variables of the cross-covariance matrix, then $\varpi
_{1}<\delta _{1}.$

The proof is evident, for TCA maximizes $\delta _{1}.$

\subsubsection{Absolute and relative contributions of a quadrant in CA}

A better known decomposition, similar to (31), where CA is interpreted as
Hotelling's canonical correlation analysis, is:

\begin{eqnarray}
\sigma _{\alpha } &=&\mathbf{v}_{\alpha }^{\prime }\ \mathbf{P}\ \mathbf{u}%
_{\alpha }  \notag \\
&=&\mathbf{v}_{\alpha }^{\prime }\ \mathbf{Q}^{(\alpha )}\ \mathbf{u}%
_{\alpha }\ \ \ \ \text{by (16),}  \notag \\
&=&\sum_{i=1}^{I}\sum_{j=1}^{J}\ v_{\alpha }(i)u_{\alpha }(j)q_{ij}^{(\alpha
)}  \notag \\
&=&\sum_{F=T,\overline{T}}\sum_{E=S,\overline{S}}\sum_{(i,j)\in S\times
T}v_{\alpha }(i)u_{\alpha }(j)q_{ij}^{(\alpha )}.
\end{eqnarray}

Equation (32) shows that, the dispersion measure $\sigma _{\alpha }=corr(%
\mathbf{v}_{\alpha },\mathbf{u}_{\alpha })$ is a weighted correlation
measure, and is decomposed into four positive parts, each part representing
the absolute contribution of a quadrant $E\times F$ to $\sigma _{\alpha }$.
From the four positive parts in (32), we define the signed absolute
(respectively relative) contribution of the quadrant $sACQ$ (respectively $%
sRCQ$) to $\sigma _{\alpha }$
\begin{equation*}
sACQ_{\alpha }(E,F)=\sum_{(i,j)\in E\times F}sign(v_{\alpha
}(i))sign(u_{\alpha }(j))v_{\alpha }(i)u_{\alpha }(j)q_{ij}^{(\alpha )},
\end{equation*}%
and%
\begin{equation*}
sRCQ_{\alpha }(E,F)=sACQ_{\alpha }(E,F)/\sigma _{\alpha },
\end{equation*}%
for $E=S$ and $\overline{S},$ and, $F=T$ and $\overline{T}.$

Similarly we define the signed residual contribution $sRES_{\alpha }$ to be
\begin{equation*}
sRES_{\alpha }=\sum_{F=T,\overline{T}}\sum_{E=S,\overline{S}}sACQ_{\alpha
}(E,F),
\end{equation*}%
which can be interpreted as a global index of attractive or repulsive
association of a CA principal dimension.

\subsubsection{Two blocks diagonal contingency tables}

Here we discuss the particular case of contingency tables which have two
blocks diagonal structure; that is, for binary partitions $I=I_{1}\cup I_{2}$
and $J=J_{1}\cup J_{2},$ $p_{ij}=0$ for $(i,j)\in I_{1}$x$J_{2\text{ }}$ and
$(i,j)\in I_{2}$x$J_{1\text{ }}.$ Then a well-known result due to Benz\'{e}%
cri (1973, p. 188-190) or Bastin et. al.(1980, pp. 174-179) is \bigskip

\textbf{Theorem 10}: (Benz\'{e}cri): A contingency table has two blocks
diagonal structure if and only if $\sigma _{1}=1;$ moreover, $%
f_{1}(i)=g_{1}(j)=c_{1}$ for $(i,j)\in I_{1}$x$J_{1\text{ }}$ and $%
f_{1}(i)=g_{1}(j)=c_{2}$ for $(i,j)\in I_{2}$x$J_{2\text{ }},$ where $c_{1}$
and $c_{2}$ are constants. \bigskip

\textbf{Corollary 11}: If $\sigma _{1}=1,$ then

a) $CA\_QSR_{1}(\mathbf{v}%
_{1+},\mathbf{u}_{1-})=CA\_QSR_{1}(\mathbf{v}_{1-},\mathbf{u}_{1+})=-1;$

b) $%
sACQ_{1}(\mathbf{v}_{1+},\mathbf{u}_{1-})=sACQ_{1}(\mathbf{v}_{1-},\mathbf{u}%
_{1+}).\bigskip $

\textbf{Remark}:

a) Benz\'{e}cri (1973, p. 188-190) generalized the result of Theorem 10 to
k-blocks diagonal contingency tables.

b) Benz\'{e}cri (1973, p.189-190) observed that it is rare to have $\sigma
_{1}=1,$ but not uncommon\ to have $\sigma _{1}^{2}\geq 0.7;$ then in these
cases the structure of the contingency table may be either \textit{quasi-two
blocks diagonal}, where few cells will be nonzero in the quadrants $I_{1}$x$%
J_{2\text{ }}$ and $I_{2}$x$J_{1\text{ }};$ or not, see Benz\'{e}cri (1973,
p.246) and Choulakian and de Tibeiro (2013).

Rodent species abundance data set, discussed in the next section provides an
example of a sparse contingency table having quasi-two blocks diagonal
structure; furthermore, it shows that $\sigma _{1}=1$ is a sufficient
condition but not necessary to have $minCA\_QSR_{1}(-)=-1.$

\subsection{Two new unified formulas}

Choulakian (2006) showed that both CA and TCA satisfy few fundamental identical
formulas -such as data reconstruction formula (14) and the transition formulas
(27, 28) of Lemma 6, even though
mathematically they are different. This paper extends the similarity of
both methods by showing that the dispersion measures also can be represented
in a common form. For $\alpha =1,...,k$%
\begin{equation*}
\varpi _{\alpha }\text{ and }\delta _{\alpha }=\sum_{(i,j)\in I\times
J}sign(f_{\alpha }(i))sign(g_{\alpha }(j))p_{i\ast }p_{\ast
j}d_{ij}^{(\alpha )},
\end{equation*}%
and%
\begin{equation*}
\sigma _{\alpha }\text{ and }\delta _{\alpha }=\sum_{(i,j)\in I\times
J}v_{\alpha }(i)u_{\alpha }(j)p_{i\ast }p_{\ast j}d_{ij}^{(\alpha )}.
\end{equation*}

\subsection{Uncomparability of CA and TCA contribution maps}

CA and TCA maps, Figures 1 and 2, are comparable because they are based on
the same data reconstruction formula (14)%
\begin{equation*}
\frac{p_{ij}}{p_{i\ast }p_{\ast j}}-1=\sum_{\alpha =1}^{k}f_{\alpha
}(i)g_{\alpha }(j)/\gamma _{\alpha },
\end{equation*}%
where $\gamma _{\alpha }^{2}=\sigma _{\alpha }^{2}=\sum_{i=1}^{I}p_{i\ast
}f_{\alpha }(i)^{2}=\sum_{j=1}^{J}p_{\ast j}g_{\alpha }(j)^{2}$ in CA and $%
\gamma _{\alpha }=\delta _{\alpha }=\sum_{i=1}^{I}p_{i\ast }|f_{\alpha
}(i)|=\sum_{j=1}^{J}p_{\ast j}|g_{\alpha }(j)|$ in TCA. Figures 1 and 2 are
obtained by plotting the coordinates ($f_{1}(i),f_{2}(i))$ and ($%
g_{1}(j),g_{2}(j)).$

Figures 3 and 4 representing TCA and CA contribution maps are not
comparable, because they do not represent the same object.

TCA contribution (TCov) is the factoring of the cross-covariance matrix by
TSVD:

\begin{equation*}
p_{ij}-p_{i\ast }p_{\ast j}=\sum_{\alpha =1}^{k}a_{\alpha }(i)b_{\alpha
}(j)/\delta _{\alpha },
\end{equation*}%
where $\delta _{\alpha }=\sum_{i=1}^{I}|a_{\alpha
}(i)|=\sum_{j=1}^{J}|b_{\alpha }(j)|$\ \ \ for $\alpha =1,...,k.$

CA contribution is the factoring of the chi-square residuals by SVD:

\begin{equation*}
\frac{p_{ij}-p_{i\ast }p_{\ast j}}{\sqrt{p_{i\ast }p_{\ast j}}}=\sum_{\alpha
=1}^{k}a_{\alpha }(i)b_{\alpha }(j)/\sigma _{\alpha },
\end{equation*}%
where $\sigma _{\alpha }^{2}=\sum_{i=1}^{I}|a_{\alpha
}(i)|^{2}=\sum_{j=1}^{J}|b_{\alpha }(j)|^{2}$\ \ \ for $\alpha =1,...,k;$
this is symmetric scaling, different from the one used by Greenacre (2013).

Figures 3 and 4 are obtained by plotting the coordinates ($a_{1}(i),a_{2}(i))
$ and ($b_{1}(j),b_{2}(j)).$ We observe that TCov map really represents TCA
contribution plot, while CA contribution map represents only the orderings
of the points according to their contributions. A really representative CA
contribution plot should be based on the coordinates $a_{\alpha }(i)^{2}\ $%
sign($a_{\alpha }(i))\ $or $b_{\alpha }(j)^{2}\ $sign($b_{\alpha }(j)).$

\section{Applications}

In this section First we revisit WS data set in detail; then consider
briefly two other data sets.

\subsection{WS brand-attribute count data}

Table 3 displays the quality of signs of residuals ($QSR$) measures in \% of
WS data: We use it to choose the number of principal dimensions. Our
interest focuses on $maxQSR_{\alpha }(+),$ because maps essentially reflect
attractive association between two optimal subsets of the row and column
categories. The first two with values of 100\% and above 89.29\% are
significant compared to the 3rd with value of 63.55\%. QSR values, $%
CA\_QSR_{2}=75.6\%$ $>QSR_{2}=72.99\%,$ show that the CA map is slightly
preferable to the TCA map. Given that CA and TCA maps Figures 1 and 2 are 
very similar, we use TCov-TCA  framework, introduced in this
essay, to interpret this data set; for there is a lot of literature on the
interpretation of CA maps,  starting with the pioneering work of Benzécri
(1973, Vol.2, chapter 2) and the few references that we cited in the introduction.
\bigskip

\begin{tabular}{l|l|l|l|l}
\multicolumn{5}{l}{\textbf{Table 3: QSR values (in \%) of WS data for the
first 4 dimensions.}} \\ \hline
$\alpha $ & $QSR_{\alpha }(+)$ & $QSR_{\alpha }(-)$ & $QSR_{\alpha }$ & $%
\delta _{\alpha }$ \\ \hline
1 & (\textbf{100,\ }52.05) & (-52.65,\ \textbf{-87.07)} & 67.01 & 0.0476 \\
2 & (65.62,\ \textbf{89.29)} & (\textbf{-84.01,\ }-60.77) & \textbf{72.99} &
0.0318 \\
3 & (62.50,\ 63.55) & (-78.88, -62.65) & 66.25 & 0.0203 \\
4 & (58.13, 64.54) & (-46.90, -80.23) & 60.17 & 0.0130 \\ \hline\hline
$\alpha $ & $CA\_QSR_{\alpha }(+)$ & $CA\_QSR_{\alpha }(-)CA\_$ & $%
CA\_QSR_{\alpha }$ & $\varpi _{\alpha }$ \\ \hline
1 & (\textbf{100,\ }52.05) & \textbf{(\ }-52.65\textbf{,-87.07)} & 67.01 &
0.0476 \\
2 & (67.95,\ \textbf{89.80)} & (\textbf{-90.24,\ }-62.60) & \textbf{75.60} &
0.0312 \\
3 & (58.20,\ 44.56) & (-49.68, -59.46) & 52.24 & 0.0155 \\
4 & (88.57, 48.59) & (-57.37, -69.26) & 62.76 & 0.0130 \\ \hline
\end{tabular}

\bigskip
Figure 3 represents taxicab interbattery analysis TCov map, which can also
be interpreted as TCA contribution map. In Figure 3, brand $i$ is
represented by the principal coordinates $(a_{1}(i),a_{2}(i)),$ and
attribute $j$ is represented by the principal coordinates $%
(b_{1}(j),b_{2}(j)).$ 

{\tiny
\begin{tabular}{l|llllllll|l}
\multicolumn{10}{l}{\textbf{Table 4: WS Covariance matrix (}$\times $\textbf{%
10}$^{4})$\textbf{\ seriated along its first TCOV principal coordinates.}}
\\ \hline
\multicolumn{10}{c}{Attribute} \\
Company & innovative & trusted & rapport & efficient & solution & leader &
relevant & essential & $\mathbf{a}_{1}(i)$ \\ \hline
Nokia & \textbf{60} & \textbf{5} & \multicolumn{1}{|l}{$-12$} & $-12$ & $-11$
& $8$ & $-12$ & $-27$ & $129$ \\
Oracle & \textbf{25} & \textbf{3} & \multicolumn{1}{|l}{$-3$} & $-1$ & $-6$
& $11$ & $-16$ & $-12$ & $57$ \\
A & \textbf{4} & \textbf{3} & \multicolumn{1}{|l}{$7$} & $5$ & $11$ & $-18$
& $-7$ & $-6$ & $15$ \\
B & \textbf{3} & \textbf{4} & \multicolumn{1}{|l}{$3$} & $3$ & $-1$ & $-6$ &
$-4$ & $-2$ & $13$ \\
E & \textbf{1} & \textbf{4} & \multicolumn{1}{|l}{$1$} & $0$ & $0$ & $-2$ & $%
-2$ & $-1$ & $10$ \\
C & \textbf{4} & \textbf{0} & \multicolumn{1}{|l}{$1$} & $1$ & $-1$ & $-1$ &
$-1$ & $-2$ & $8$ \\
G & \textbf{1} & \textbf{2} & \multicolumn{1}{|l}{$2$} & $-1$ & $-1$ & $-3$
& $-1$ & $1$ & $6$ \\ \cline{2-9}
D & $0$ & $-1$ & \multicolumn{1}{|l}{$3$} & $0$ & $3$ & $-5$ & $0$ & $0$ & $%
-2$ \\
I & $-13$ & $1$ & \multicolumn{1}{|l}{$-10$} & $-13$ & $8$ & $10$ & $-2$ & $%
\mathbf{20}$ & $-25$ \\
F & $-37$ & $8$ & \multicolumn{1}{|l}{$6$} & $2$ & $\mathbf{16}$ & $-6$ & $%
\mathbf{12}$ & $0$ & $-58$ \\
H & $-17$ & $-13$ & \multicolumn{1}{|l}{$1$} & $6$ & $-4$ & $3$ & $\mathbf{17%
}$ & $8$ & $-60$ \\
Fedex & $-30$ & $-16$ & \multicolumn{1}{|l}{$2$} & $11$ & $-14$ & $9$ & $%
\mathbf{17}$ & $\mathbf{20}$ & $-93$ \\ \hline
$\mathbf{b}_{1}(j)$ & $194$ & $44$ & $-4$ & $-10$ & $-18$ & $-23$ & $-86$ & $%
-98$ &  \\ \hline
\end{tabular}%
}

\subsubsection{Interpretation of the first principal dimension of TCov map}

Table 4 displays the cross-covariance matrix seriated
along its first TCov principal coordinates; where we clearly also observe the
four principal quadrants for $\alpha =1.$
 
Let  $T=\left\{ innovative, trusted \right\} $ and 
 $S=\left\{Nokia, Oracle, A, B, C, G \right\} $ for the first principal dimension. 
 In Table 4 we observe that the covariance values in $S\times T$ quadrant are
all positive; and that is the reason that max$QSR_{1}(+)=100\%,$ the highest
attainable value. Similarly one observes that for the first two principal
dimensions $minQSR_{\alpha }(-)$ are $-87.07\%$ and $-84.04\%$, relatively
significant values compared to the lower bound $-100\%$. The first taxicab
dispersion measure, displayed in Table 3, is $\delta _{1}=0.0476=%
\displaystyle\sum\limits_{i=1}^{12}|a_{1}(i)|=\displaystyle%
\sum\limits_{i=1}^{8}|b_{1}(j)|$ by (18)$.$ The last column and the last row
of Table 4 display the signed absolute contributions $a_{1}(i)$ and $b_{1}(j)
$ to $\delta _{1}$, that according to Benz\'{e}cri (1973, p.47) assist in
the interpretation of the first factor. So the relative contribution (RC) of
\textit{innovative} to the first factor is $RC_{1}(innovative)=194/476=0.41%
\leq 0.5$, a very high value indeed. Similarly the $RC_{1}(relevant$ or $%
essential)=(98+86)/476=0.39\leq 0.5$, a very high value. In $TCov(\mathbf{P})
$, the attainable upper bound of a RC of a coordinate $a_{\alpha }(i)$ or $%
b_{\alpha }(j)$ is 0.5 by equation (22). So the first principal dimension
represents the factor opposing (\textit{innovative} associated with \textit{%
Nokia, and }with lesser degree with \textit{Oracle}) to (\textit{%
relevant-essential} associated with the brands \textit{Fedex, H and F}).
Furthermore, the cross- covariance matrix in Table 4 informs us more about
this opposition: \textit{cov(Nokia, innovative)=60 }and\textit{\ cov(Oracle,
innovative)=25, }while \textit{cov(F, innovative)=-37, cov(Fedex,
innovative)=-30, }and \textit{cov(H, innovative)=-17.} Note that in
particular the intensity of the negative covariances, representing the three
major repulsive associations, can not be assessed in Figure 2. So, to assess
quantitatively an association between a row and a column, one has to follow
Collins advice and look at the value in a table of numbers.

Furthermore, examining the seriated Table 2, we see that the last three
brands \textit{Fedex, H and F }do not have any molehills: they satisfy
equation (25) of Lemma 4, and they have quite large marginal weights.

\subsubsection{Interpretation of the second principal dimension of TCov map}

Table 5 displays the residual cross-covariance matrix \textbf{P}$^{(2)}$
seriated along its 2nd TCov principal coordinates: It shows that the second
principal or latent variable is based on the opposition between (\textit{%
leader-innovative} associated essentially with \textit{Nokia}) and (\textit{%
solution-rapport-trusted} associated with brands \textit{F and A}).

Figure 2 shows that each of the brands $\left\{ C,G,D,E,B\right\} $ have
very small almost insignificant contributions either to the first or to the
second principal dimensions; and this is also evident in Tables 4 and 5.

\bigskip

{\tiny \
\begin{tabular}{l|llllllll|l}
\multicolumn{10}{l}{\textbf{Table 5: WS P}$^{(2)}$\textbf{\ matrix (}$\times
$\textbf{10}$^{4})$\textbf{\ seriated along its second TCOV principal
coordinates.}} \\ \hline
\multicolumn{10}{c}{Attribute} \\
Company & leader & innovative & essential & relevant & efficient & trusted &
rapport & solution & $\mathbf{a}_{2}(i)$ \\ \hline
Nokia & \textbf{14} & \textbf{7} & $-0$ & $12$ & \multicolumn{1}{|l}{$-9$} &
-7 & -11 & -6 & 65 \\
fedex & \textbf{5} & 8 & $-1$ & $+0$ & \multicolumn{1}{|l}{$-9$} & -8 & 2 &
-17 & 29 \\
I & 9 & -3 & 15 & -7 & \multicolumn{1}{|l}{-14} & 3 & -10 & 7 & 28 \\
H & -0 & 8 & -4 & 6 & \multicolumn{1}{|l}{5} & -8 & 0 & -7 & 19 \\
Oracle & 14 & 2 & -0 & -6 & \multicolumn{1}{|l}{0} & -2 & -3 & -4 & 17 \\
\cline{2-9}
C & -1 & 0 & -1 & 1 & \multicolumn{1}{|l}{1} & -0 & 1 & -1 & -1 \\
G & -3 & -2 & 2 & 1 & \multicolumn{1}{|l}{-1} & 2 & 2 & -1 & -4 \\
E & -2 & -3 & 1 & -1 & \multicolumn{1}{|l}{0} & 3 & 1 & 0 & -9 \\
D & -5 & 1 & -1 & -0 & \multicolumn{1}{|l}{-0} & -1 & 3 & 3 & -10 \\
B & -5 & -2 & 1 & -2 & \multicolumn{1}{|l}{4} & 2 & 3 & -1 & -18 \\
A & -17 & -2 & -3 & -4 & \multicolumn{1}{|l}{5} & 2 & 7 & 12 & -52 \\
F & -9 & -13 & -12 & 1 & \multicolumn{1}{|l}{0} & 13 & 6 & 14 & -66 \\ \hline
$\mathbf{b}_{2}(j)$ & 83 & 42 & 24 & 10 & -18 & -42 & -45 & -53 &  \\ \hline
\end{tabular}%
}

\subsubsection{TCA map}

Equation (17) shows that the TCA map, Figure 2, is a change of scale of the
TCov map, Figure 3. In Figure 2, brand $i$ is represented by the principal
coordinates $(f_{1}(i),f_{2}(i)),$ and attribute $j$ is represented by the
principal coordinates $(g_{1}(j),g_{2}(j)).$ The first row and column
principal coordinates are given in Table 6 in decreasing order, and
accordingly the $(density-1=\frac{p_{ij}}{p_{i\ast }p_{\ast j}}-1$) values
of the attribute and brand categories are displayed in Table 6. However, the
relative position of some brands and attributes in Figure 2 are completely
different in Figure 3. For instance, in TCA map Figure 2 on dimension 1 the
\textit{brand C} seems to be much more important than the brands \textit{%
Oracle or Nokia}: By Lemma 6, the covariance of \textit{brand C} with the
first factor is $f_{1}(C)=13/100,$ which is much larger than $%
f_{1}(Nokia)=9.3/100$; while in TCov map Figure 3 it is the opposite, the
contributions are $a_{1}(C)=8/10000$ and $a_{1}(Nokia)=129/10000$. This
aspect is the cause of confusion and difficulty in the interpretation of CA
or TCA maps. Lemma 6 helps us to explain this fact: The \textit{brand }$C$
is \textit{strongly associated} with the latent variable \textit{innovative,
but it does not contribute to the construction of this latent variable. }%
While \textit{Nokia} is \textit{moderately associated} with the latent
variable \textit{innovative, even though it constructs this latent variable;
because it also constructs the 2nd principal dimension. } That is, the TCov
map helps us to identify rows or columns that essentially contribute to the
formation of a latent variable, while the TCA map shows us the rows and the
columns which are highly associated with the latent variables defined by the TCov decomposition.
\bigskip

{\tiny
\begin{tabular}{l|llllllll|l}
\multicolumn{10}{l}{\textbf{Table 6: WS (density -1) matrix (}$\times $%
\textbf{\ }$100)$\textbf{\ seriated along the first TCA principal dimension.}
} \\ \hline
\multicolumn{10}{c}{Attribute} \\
Company & innovative & trusted & rapport & efficient & solution & leader &
relevant & essential & $f_{1}(i)$ \\ \hline
C & \textbf{54} & \textbf{4} & \multicolumn{1}{|l}{$9$} & $6$ & $-13$ & $-17$
& $-9$ & $-67$ & $13$ \\
Oracle & \textbf{40} & \textbf{4} & \multicolumn{1}{|l}{$-4$} & $-1$ & $-9$
& $14$ & $-23$ & $-38$ & $10$ \\
B & \textbf{21} & \textbf{17} & \multicolumn{1}{|l}{$19$} & $17$ & $-7$ & $%
-32$ & $-28$ & $-29$ & $10$ \\
E & \textbf{7} & \textbf{24} & \multicolumn{1}{|l}{$6$} & $0$ & $-1$ & $-18$
& $-19$ & $-10$ & $9.4$ \\
Nokia & \textbf{39} & \textbf{2} & \multicolumn{1}{|l}{$-6$} & $-6$ & $-6$ &
$4$ & $-7$ & $-34$ & $9.3$ \\
G & \textbf{7} & \textbf{16} & \multicolumn{1}{|l}{$18$} & $-10$ & $-12$ & $%
-24$ & $-5$ & $16$ & $6.8$ \\
A & \textbf{10} & \textbf{5} & \multicolumn{1}{|l}{$12$} & $8$ & $24$ & $-32$
& $-14$ & $-25$ & $3.8$ \\ \cline{2-9}
D & $1$ & $-5$ & \multicolumn{1}{|l}{$15$} & $-1$ & $22$ & $-26$ & $0$ & $-6$
& $-1.5$ \\
H & $-7$ & $-4$ & \multicolumn{1}{|l}{$0$} & $2$ & $-2$ & $1$ & $6$ & $7$ & $%
-2.7$ \\
I & $-14$ & $1$ & \multicolumn{1}{|l}{$-9$} & $-10$ & $7$ & $9$ & $-2$ & $42$
& $-2.9$ \\
Fedex & $-10$ & $-4$ & \multicolumn{1}{|l}{$1$} & $3$ & $-4$ & $3$ & $5$ & $%
13$ & $-3.4$ \\
F & $-25$ & $4$ & \multicolumn{1}{|l}{$3$} & $1$ & $10$ & $-3$ & $7$ & $0$ &
$-4.5$ \\ \hline
$g_{1}(j)$ & $17.6$ & $2.7$ & $-0.3$ & $-0.7$ & $-1.5$ & $-1.6$ & $-6.9$ & $%
-17.3$ &  \\ \hline
\end{tabular}
}

\subsubsection{WS data set by CA}

Table 3 displays the $CA\_QSR$ values for the WS data set: $CA\_QSR_{1}$ and
$QSR_{1}$ values are identical by Lemma 10; $CA\_QSR_{2}$ values being a
little bit better than the corresponding $QSR_{2}$ values. This is the main
reason that both CA and TCA maps, Figures 1 and 2, are very similar.

Table 7 represents the $sACQ$ and $sRSQ$ values for the WS data set, which
reflect, somewhat in a different way, the $CA\_QSR$ values. For the first
principal dimension, there is a positively associated quadrant which
contributes $44.15\%$ to $\sigma _{1}$, and a negatively associated quadrant
whose contribution to $\sigma _{1}$ is $27.89\%.$ Furthermore, we note that
in the first principal dimension globally the excess attractive association
is quite small $100sRES_{1}/\sigma _{1}=9.9\%.$ For the second principal
dimension, there is a negatively associated quadrant which contributes $%
48.43\%$ to $\sigma _{2}$; a postively associated quadrant whose
contribution to $\sigma _{2}$ is $28.43\%;$ and a small excess repulsive
association $100sRES_{2}/\sigma _{2}=-12.14\%$.

\bigskip
\begin{tabular}{l|l|l|l|l}
\multicolumn{5}{l}{\textbf{Table 7: sACQ and sRCQ of WS data for the first 4
dimensions of CA.}} \\ \hline
$\alpha $ & $100sACQ_{\alpha }(+)$ & $100sACQ_{\alpha }(-)$ & $%
100sRES_{\alpha }$ & $100\sigma _{\alpha }$ \\ \hline
1 & (\textbf{4.02,\ }0.98) & (-1.56,\ \textbf{-2.54)} & 0.90 & 9.10 \\
2 & (0.76,\ \textbf{1.39)} & (\textbf{-2.38,\ }-0.37) & -0.59 & 4.90 \\
3 & (1.13,\ 0.29) & (-0.29, -1.02) & 0.11 & 2.72 \\
4 & (0.48, 0.42) & (-0.52, -0.73) & -0.35 & 2.15 \\ \hline\hline
$\alpha $ & $100sRCQ_{\alpha }(+)$ & $100sRCQ_{\alpha }(-)$ & $%
100sRES_{\alpha }/\sigma _{\alpha }$ & $100\sigma _{\alpha }/\sigma _{\alpha
}$ \\ \hline
1 & (\textbf{44.15,\ }10.80) & \textbf{(\ }-17.16\textbf{,-27.89)} & 9.90 &
100 \\
2 & (15.50,\ \textbf{28.43)} & (\textbf{-48.43,\ }-7.46) & -12.14 & 100 \\
3 & (41.48,\ 10.50) & (-10.69, -37.34) & 3.95 & 100 \\
4 & (22.39, 19.38) & (-24.25, -33.98) & -16.47 & 100 \\ \hline
\end{tabular}

\bigskip

Here, we also want to remind the reader that CA and TCA of two other data
sets of marketing research, discussed in Bock (2011a) and Bendixen (1996),
produced similar results to the analysis of WS data set by both methods.

\subsection{Faust data set}

Faust (2005) analyzed by CA a two-mode affiliation network, (0-1) matrix $%
\mathbf{Z=(}z_{ij})$ of size $22\times 15$; where the 22 rows represent 22
countries and the 15 columns the regional trade and treaty organizations in
the American continent. The country $i$ is a member of the organization $j$
if $z_{ij}=1$; and $z_{ij}=0$ means the country $i$ is not a member of the
organization $j$. Choulakian and Abou-Samra (2020) compared the CA and TCA
maps, where they found that the TCA map is much more interpretable than the
corresponding CA map because of the existence of some influential columns
and rows (outliers?) that dominated the second CA principal dimension. Table
8 displays the QSR measures for both methods. $QSR_{1}=63.67\%$ is of
comparable value to $CA\_QSR_{1}=61.25\%.$ However the corresponding values
for the second dimension are completely different: $QSR_{2}=57.96\%$ and $%
CA\_QSR_{2}=33.41\%,$ a significant difference of $24.55\%$. This is also
reflected in the $sRCQ_{2}$ values reported in Table 9, where a positively
associated quadrant contributes $78.56\%$ to $\sigma _{2}=0.5331$; while in
TCA this value is constant and equals 25\%. Furthermore, $100sRES_{2}/\sigma
_{2}=63.41\%$ represents a significant excess of attractive
association.\bigskip\

\begin{tabular}{l|l|l|l|l}
\multicolumn{5}{l}{\textbf{Table 8: QSR values (in \%) of Faust data for the
first 4 dimensions.}} \\ \hline
$\alpha $ & $QSR_{\alpha }(+)$ & $QSR_{\alpha }(-)$ & $QSR_{\alpha }$ & $%
\delta _{\alpha }$ \\ \hline
1 & (\textbf{56.52,\ }53.79) & (-67.43,\ \textbf{-85.43)} & 63.67 & 0.4266
\\
2 & (43.78,\ \textbf{70.80)} & (\textbf{-68.79,\ }-57.12) & \textbf{57.96} &
\textbf{0.2751} \\
3 & (65.51,\ 44.38) & (-78.00, -82.57) & 63.77 & 0.2659 \\
4 & (61.45, 55.04) & (-67.63, -44.33) & 55.72 & 0.1782 \\ \hline\hline
$\alpha $ & $CA\_QSR_{\alpha }(+)$ & $CA\_QSR_{\alpha }(-)CA\_$ & $%
CA\_QSR_{\alpha }$ & $\varpi _{\alpha }$ \\ \hline
1 & (\textbf{69.71,\ }44.86) & \textbf{(\ }-59.20\textbf{,-84.94)} & 61.25 &
0.4104 \\
2 & (17.51,\ \textbf{63.72)} & (\textbf{-49.87,\ }-37.22) & \textbf{33.41} &
\textbf{0.1570} \\
3 & (60.29,\ 43.54) & (-81.95, -45.06) & 54.09 & 0.2195 \\
4 & (33.26, 55.77) & (-46.57, -54.61) & 45.57 & 0.1636 \\ \hline
\end{tabular}%
\bigskip

\bigskip

\bigskip
\begin{tabular}{l|l|l|l|l}
\multicolumn{5}{l}{\textbf{Table 9: sACQ and sRCQ of Faust data for the
first 4 dimensions of CA.}} \\ \hline
$\alpha $ & $100sACQ_{\alpha }(+)$ & $100sACQ_{\alpha }(-)$ & $%
100sRES_{\alpha }$ & $100\sigma _{\alpha }$ \\ \hline
1 & (\textbf{22.56,\ }10.21) & (-11.13,\ \textbf{-15.90)} & 5.74 & 59.80 \\
2 & (1.68,\ \textbf{41.89)} & (\textbf{-5.75,\ }-4.00) & 33.81 & 53.31 \\
3 & (4.55,\ 25.96) & (-8.26, -6.22) & 16.02 & 44.99 \\
4 & (8.32, 11.90) & (-11.00, -8.17) & 1.06 & 39.40 \\ \hline\hline
$\alpha $ & $100sRCQ_{\alpha }(+)$ & $100sRCQ_{\alpha }(-)$ & $%
100sRES_{\alpha }/\sigma _{\alpha }$ & $100\sigma _{\alpha }/\sigma _{\alpha
}$ \\ \hline
1 & (\textbf{37.72,\ }17.08) & \textbf{(\ }-18.62\textbf{,-26.58)} & 9.60 &
100 \\
2 & (3.14,\ \textbf{78.56)} & (\textbf{-10.79,\ }-7.50) & \textbf{63.41} &
100 \\
3 & (10.11,\ 57.70) & (-18.36, -13.83) & 35.62 & 100 \\
4 & (21.13, 30.21) & (-27.92, -20.74) & 2.68 & 100 \\ \hline
\end{tabular}

\subsection{Rodent species abundance data}

Table 10 displays the seriated Rodent species abundance data of size $28
\times\ 9$, where 9 species of rodents have been counted at each of 28 sites
in California. The data set is from Quinn and Keough (2002), but is
available in the R package TaxicabCA. Choulakian (2017) presented a detailed
analysis of this data set by CA and TCA. Given that, $\sigma _{1}^{2}\geq
0.7465,$ a value greater than 0.7 marked by
Benz\'{e}cri, we clearly observe
the quasi-two blocks diagonal structure in Table 10.

{\tiny
\begin{tabular}{lllllllllll}
\multicolumn{10}{l}{\textbf{Table 10: CA seriated rodent species abundance
data.}} &  \\ \hline
\textit{Sites} & \textit{rod1} & \textit{rod2} & \textit{rod6} & \textit{rod8%
} & \textit{rod3} & \textit{rod5} & \textit{rod9} & \textit{rod4} & \textit{%
rod7} & \multicolumn{1}{||l}{$f_{1}(site)$} \\ \hline
\textit{24} & \multicolumn{1}{|l}{\textbf{1}} &  & \multicolumn{1}{|l}{} &
&  &  &  &  &  & \multicolumn{1}{||l}{3.49} \\
\textit{17} & \multicolumn{1}{|l}{\textbf{3}} &  & \multicolumn{1}{|l}{} &
&  &  &  &  &  & \multicolumn{1}{||l}{3.49} \\
\textit{21} & \multicolumn{1}{|l}{\textbf{2}} & \textbf{1} &
\multicolumn{1}{|l}{} &  &  &  &  &  &  & \multicolumn{1}{||l}{3.35} \\
\textit{10} & \multicolumn{1}{|l}{\textbf{1}} & \textbf{2} &
\multicolumn{1}{|l}{} &  &  &  &  &  &  & \multicolumn{1}{||l}{3.21} \\
\textit{9} & \multicolumn{1}{|l}{\textbf{3}} & \textbf{8} &
\multicolumn{1}{|l}{} &  &  &  &  &  &  & \multicolumn{1}{||l}{3.18} \\
\textit{14} & \multicolumn{1}{|l}{\textbf{1}} & \textbf{3} &
\multicolumn{1}{|l}{} &  &  &  &  &  &  & \multicolumn{1}{||l}{3.17} \\
\textit{7} & \multicolumn{1}{|l}{} & \textbf{11} & \multicolumn{1}{|l}{} &
&  &  &  &  &  & \multicolumn{1}{||l}{3.07} \\
\textit{11} & \multicolumn{1}{|l}{} & \textbf{9} & \multicolumn{1}{|l}{} &
&  &  &  &  &  & \multicolumn{1}{||l}{3.07} \\
\textit{15} & \multicolumn{1}{|l}{} & \textbf{11} & \multicolumn{1}{|l}{} &
&  &  &  &  &  & \multicolumn{1}{||l}{3.07} \\
\textit{25} & \multicolumn{1}{|l}{} & \textbf{5} & \multicolumn{1}{|l}{} &
&  &  &  &  &  & \multicolumn{1}{||l}{3.07} \\
\textit{22} & \multicolumn{1}{|l}{} & \textbf{3} & \multicolumn{1}{|l}{} &
&  &  &  &  &  & \multicolumn{1}{||l}{3.07} \\
\textit{8} & \multicolumn{1}{|l}{} & \textbf{16} &  &  &  &  &  &  &  &
\multicolumn{1}{||l}{3.07} \\
\textit{16} & \multicolumn{1}{|l}{} & \textbf{4} & \multicolumn{1}{|l}{} &
&  &  &  &  &  & \multicolumn{1}{||l}{3.07} \\
\textit{1} & \multicolumn{1}{|l}{} & \textbf{13} & \multicolumn{1}{|l}{2} &
& 3 & 1 &  & 1 &  & \multicolumn{1}{||l}{1.85} \\ \cline{2-11}
\textit{20} & \multicolumn{1}{|l}{3} &  & \multicolumn{1}{|l}{} &  & \textbf{%
27} &  &  & \textbf{1} &  & \multicolumn{1}{||l}{-0.4} \\
\textit{12} & \multicolumn{1}{|l}{} & 3 & \multicolumn{1}{|l}{\textbf{16}} &
\textbf{7} & \textbf{1} & \textbf{5} &  &  &  & \multicolumn{1}{||l}{-0.5}
\\
\textit{3} & \multicolumn{1}{|l}{} & 4 & \multicolumn{1}{|l}{\textbf{9}} &
& \textbf{36} & \textbf{2} &  &  &  & \multicolumn{1}{||l}{-0.13} \\
\textit{13} & \multicolumn{1}{|l}{} & 4 & \multicolumn{1}{|l}{\textbf{12}} &
& \textbf{39} & \textbf{4} &  &  &  & \multicolumn{1}{||l}{-0.17} \\
\textit{4} & \multicolumn{1}{|l}{} & 4 & \multicolumn{1}{|l}{\textbf{30}} &
\textbf{18} & \textbf{53} & \textbf{5} & \textbf{3} & \textbf{1} &  &
\multicolumn{1}{||l}{-0.28} \\
\textit{18} & \multicolumn{1}{|l}{} & 2 & \multicolumn{1}{|l}{\textbf{14}} &
\textbf{4} & \textbf{78} & \textbf{10} &  &  &  & \multicolumn{1}{||l}{-0.34}
\\
\textit{5} & \multicolumn{1}{|l}{} & 2 & \multicolumn{1}{|l}{\textbf{16}} &
& \textbf{63} & \textbf{11} &  & \textbf{21} &  & \multicolumn{1}{||l}{-0.37}
\\
\textit{23} & \multicolumn{1}{|l}{} &  & \multicolumn{1}{|l}{\textbf{8}} &
\textbf{2} &  & \textbf{2} &  &  &  & \multicolumn{1}{||l}{-0.38} \\
\textit{26} & \multicolumn{1}{|l}{} &  & \multicolumn{1}{|l}{\textbf{11}} &
\textbf{2} & \textbf{22} &  &  &  &  & \multicolumn{1}{||l}{-0.39} \\
\textit{27} & \multicolumn{1}{|l}{} &  & \multicolumn{1}{|l}{\textbf{9}} &
\textbf{1} & \textbf{29} & \textbf{10} &  &  &  & \multicolumn{1}{||l}{-0.40}
\\
\textit{19} & \multicolumn{1}{|l}{} &  & \multicolumn{1}{|l}{} &  & \textbf{1%
} &  &  &  &  & \multicolumn{1}{||l}{-0.41} \\
\textit{28} & \multicolumn{1}{|l}{} &  & \multicolumn{1}{|l}{\textbf{1}} &
& \textbf{10} &  &  & \textbf{1} &  & \multicolumn{1}{||l}{-0.42} \\
\textit{6} & \multicolumn{1}{|l}{} & 1 & \multicolumn{1}{|l}{\textbf{8}} &
\textbf{2} & \textbf{48} & \textbf{12} & \textbf{2} & \textbf{35} & \textbf{%
12} & \multicolumn{1}{||l}{-0.44} \\
\textit{2} & \multicolumn{1}{|l}{} & 1 & \multicolumn{1}{|l}{\textbf{16}} &
\textbf{2} & \textbf{57} & \textbf{9} & \textbf{3} & \textbf{65} & \textbf{8}
& \multicolumn{1}{||l}{-0.45} \\ \hline\hline
\multicolumn{1}{|l}{$g_{1}(rod)$} & \multicolumn{1}{|l}{3.02} & 2.65 &
\multicolumn{1}{|l}{-0.32} & -0.32 & -0.36 & -0.37 & -0.44 & -0.48 & -0.52 &
\multicolumn{1}{|l|}{} \\ \hline
\end{tabular}
}
\bigskip

Furthermore,

a) $CA\_QSR_{1}(\mathbf{v}_{1+},\mathbf{u}_{1+})=0.965,$ $CA\_QSR_{1}(%
\mathbf{v}_{1-},\mathbf{u}_{1-})=0.190,$

$CA\_QSR_{1}(\mathbf{v}_{1-},\mathbf{u}_{1+})=-0.943,$ $CA\_QSR_{1}(\mathbf{v%
}_{1+},\mathbf{u}_{1-})=-1.$

This shows that the quality of signs in the 4th quadrant of the $Cov(%
\mathbf{P})$( not shown), is very poor, 0.190 ; and the result in Corollary 11a is approximately
satisfied, -1 and -0.943.

b) $sACQ_{1}(\mathbf{v}_{\alpha +},\mathbf{u}_{\alpha +})=0.674,\ sACQ_{1}(%
\mathbf{v}_{1-},\mathbf{u}_{1-})=0.014,$

$sACQ_{1}(\mathbf{v}_{1-},\mathbf{u}_{1+})=-0.084,\ sACQ_{1}(\mathbf{v}_{1+},%
\mathbf{u}_{1-})=-0.092.$

This shows that the result in Corollary 11b is approximately satisfied.

c) $sRCQ_{1}(\mathbf{v}_{\alpha +},\mathbf{u}_{\alpha +})=0.7798,\ sRCQ_{1}(%
\mathbf{v}_{1-},\mathbf{u}_{1-})=0.0158,$

$sRCQ_{1}(\mathbf{v}_{1-},\mathbf{u}_{1+})=-0.0974,\ sRCQ_{1}(\mathbf{v}%
_{1+},\mathbf{u}_{1-})=-0.1070.$

This shows that 78\% of the contribution to $\sigma _{1}=0.864$ comes from
the 2nd quadrant of Table 10; a very small proportion 1.58\% comes from the
4th quadrant of Table 10; and about 20\% comes from the quasi-sparse blocks.

\section{Conclusion}

A crucial first step in data analysis of multivariate tables is the
preprocessing step: centering and/or scaling of the data. In the case of CA
of contingency tables, the row and column marginals are intricate part of
the centering and scaling of the method via the chi-square residuals defined
for the row and column profiles. In a pioneering work, Goodman (1991) and
his discussants compared the effects of marginal weighted scores and uniform
weighted scores in association and correlation models in the analysis of
contingency tables. A parallel to this problem is: Should we decompose the
covariance matrix (do Tucker interbattery analysis) or the correlation
matrix (do Hotelling canonical correlation analysis)?

This essay attempted to clarify mainly the following issues:

First, we showed that the aims of CA and TCA are different, but interrelated. The aim of CA is
to explain the heterogeneity of the row and column profiles (row and column
conditional distributions), from which as a byproduct we get a view of the
dependence structure of the cross-covariance matrix. While, the aim of TCA
is to explain the dependence structure of the cross-covariance matrix, then
as a by product obtain a view of the heterogeneity of the row and column
profiles. Empirical data have shown that the cross-covariance matrix ($%
p_{ij}-p_{i\ast }p_{\ast j})$ is much more robust than the chi-square
residual matrix ($\frac{p_{ij}-p_{i\ast }p_{\ast j}}{\sqrt{p_{i\ast }p_{\ast
j}}})$, a fact first observed by Tenenhaus and Augendre (1996).

Second, two maps are needed to fully picture the dependence/heterogeneity
structure in a contingency table; and Lemma 6 explains in a simple way the
relationship between the two maps.

Third, for a principal dimension we introduced the new concept of the intrinsic
quality and distinguished it from the often used extrinsic quality;
 and related the intrinsic quality to the quality of
signs of the residuals in the four quadrants. Furthermore, we provided
quantifications of the intrinsic quality by introducing $QSR$ and $sACQ$
indices.

Fourth, we emphasized the importance of looking at the residual covariance
values at each iteration, a general procedure exemplified by Tukey (1977) in
exploratory data analysis.

\bigskip
\begin{verbatim}
Acknowledgements
\end{verbatim}

Choulakian's research has been supported by NSERC of Canada. Choulakian
thanks Eric Beh for discussing this topic and furnishing few references; and
Ahc\`{e}ne Brahmi for help in editing.\bigskip

\textbf{References}

Allard, J., Champigny, S., Choulakian, V. and Mahdi, S. (2020). TCA and
TLRA: A comparison on contingency tables and compositional data.
https: //arxiv.org/abs/2009.05482

Aristotle (1960). \textit{The Pocket Aristotle}. Edited by Kaplan J.D, N.Y:
The Pocket Books.

Bastin, Ch., Benz\'{e}cri, J.P., Bourgarit, Ch, Cazes, P. (1980). \textit{%
Pratique de l'Analyse Des Donn\'{e}es: Vol. 2, Abr\'{e}g\'{e} Th\'{e}orique,
\'{E}tudes de Cas Mod\`{e}le}. Dunod, Paris.

Beh, E. and Lombardo, R. (2014). \textit{Correspondence Analysis: Theory,
Practice and New Strategies}. N.Y: Wiley.

Bendixen, M. (1996). A practicalguide to the use of correspondence analysis
in marketing research. http://marketing-bulletin.massey.ac.nz

Benz\'{e}cri, J.P. (1973).\ \textit{L'Analyse des Donn\'{e}es: Vol. 1: La Taxinomie; Vol. 2:
L'Analyse des Correspondances}. Paris: Dunod.

Benz\'{e}cri, J.P. (1982). Qualit\'{e} et quantit\'{e}: La grandeur et
l'espace selon Bergson et en analyse des donn\'{e}es. \textit{Les Cahiers de
l'Analyse des Donn\'{e}es}, 7(4), 395-412.

Benz\'{e}cri, J.P. (1988). Qualit\'{e} et quantit\'{e} dans la tradition des
philosophes et en analyse des donn\'{e}es. \textit{Les Cahiers de
l'Analyse des Donn\'{e}es}, 13(1), 131-152.

Bock, T. (2011a). Improving the display of correspondence analysis using moon
plots. \textit{International Journal of Market Research}, 53, 307--326.

Bock, T. (2011b). We really do need correspondence analysis. \textit{%
International Journal of Market Research}, 53, 587--591.

Bock, T. (2017). How to interpret correspondence analysis plots (It probably
isn't the way you think). Available at

\textit{%
https://www.r-bloggers.com/2017/05/how-to-interpret-correspondence-analysis-plots-it-probably-isnt-the-way-you-%
}think/

Choulakian, V. (2006). Taxicab correspondence analysis. \textit{%
Psychometrika,} 71, 333-345.

Choulakian, V. and de Tibeiro, J. (2013). Graph partitioning by
correspondence analysis and taxicab correspondence analysis. \textit{Journal
of Classification, }30, 397-427.

Choulakian, V. (2014). Taxicab correspondence analysis of ratings and
rankings. \textit{Journal de la Soci\'{e}t\'{e} Fran\c{c}aise de Statistique, }155(
4), 1-23.

Choulakian, V., Simonetti, B. and Gia, T.P. (2014). Some further aspects of
taxicab correspondence analysis. \textit{Statistical Methods and Applications%
}, 23, 401-416.

Choulakian, V. (2016a). Matrix factorizations based on induced norms.
\textit{Statistics, Optimization and Information Computing}, 4, 1-14.

Choulakian, V. (2016b). Globally homogenous mixture components and local
heterogeneity of rank data. https://arxiv.org/pdf/1608.05058.pdf

Choulakian, V. (2017). Taxicab correspondence analysis of sparse contingency
tables. \textit{Italian Journal of Applied Statistics,} 29 (2-3), 153-179.

Choulakian, V. and Abou-Samra, G. (2020). Mean absolute deviations about the
mean, the cut norm and taxicab correspondence analysis. \textit{Open Journal
of Statistics}, 10(1), 97-112.

Collins, M. (2002). Analyzing brand image data. \textit{Marketing Research},
14, 32--36.

Collins, M. (2011). Do we really need correspondence analysis? \textit{%
International Journal of Market Research}, 53, 583--586.

De Leeuw, J. (2005). Book review 5: Correspondence Analysis and Data Coding
with Java and R. \textit{Journal of Statistical Software}, 14, 1-2.

Faust, K. (2005). Using correspondence analysis for joint displays of 
affiliation networks. In: Carrington, P.J., Scott, J., Wasserman, S. (Eds.), \textit{Models
and Methods in Social Network Analysis}. Cambridge University Press, Cambridge, 117–147. 

Friendly, M. and Kwan, K. (2011). Comment (Graph people versus Table
people). \textit{Journal of Computational and Graphical Statistics}, 20 (1),
18--27.

Goodman, L.A. (1991). Measures, models, and graphical displays in the
analysis of cross-classified data. \textit{Journal of the American
Statistical Association}, 86 (4), 1085-1111.

Goodman, L.A. (1996). A single general method for the analysis of
cross-classified data: Reconciliation and synthesis of some methods of
Pearson, Yule, and Fisher, and also some methods of correspondence analysis
and association analysis.\textit{\ Journal of the American Statistical
Association}, 91, 408-428.

Greenacre, M. (2013). Contribution biplots. \textit{Journal of Computational
and Graphical Statistics}, 22(1), 107--122.

Greenacre, M. and Lewi, P. (2009). Distributional equivalence and
subcompositional coherence in the analysis of compositional data,
contingency tables and ratio-scale measurements. \textit{Journal of
Classification}, 26, 29-54.

Hotelling, H. (1933). Analysis of a complex of statistical variables into
principal components. \textit{Journal of Educational Psychology}, 24,
417--441, and 498--520.

Hotelling, H. (1936). Relations between two sets of variates. \textit{%
Biometrika}, 28(3/4), 321--377.

Lancaster, H. (1958). The structure of bivariate distributions. \textit{The
Annals of Mathematical Statistics}, 29(3), 719--736.

Murtagh F. (2005). \textit{Correspondence Analysis and Data Coding with R
and Java}. Chapman and Hall/CRC Press.

Quinn, G. and Keough, M. (2002). \textit{Experimental Design and Data
Analysis for Biologists}. Cambridge Univ. Press, Cambridge, UK.

Tenenhaus, M. and Augendre, H. (1996). Analyse factorielle inter-batteries
de Tucker et analyse canonique aux moindres carr\'{e}s partiels. In \textit{%
Recueil des r\'{e}sum\'{e}s des communications des 28\`{e}me Journ\'{e}es de
statistique}, 693-697.

Tucker, L. R. (1958). An inter-battery method of factor analysis. \textit{%
Psychometrika}, 23, 111--136.

Tukey, J.W. (1977). \textit{Exploratory Data Analysis}. Addison-Wesley:
Reading, Massachusetts.

Whitlark, D.B. and Smith, S.M. (2001). Using correspondence analysis to map
relationships. \textit{Marketing Research}, 13 (3), 22--27.

Whitlark, D.B. and Smith, S.M. (2002). Why go beyond the basics. \textit{%
Marketing Research}, 14 (3), 41.
\begin{verbatim}

\end{verbatim}

\end{document}